\documentclass[useAMS,usenatbib,a4paper]{mnras}
\usepackage{graphicx}
\usepackage{amssymb}
\usepackage{epstopdf}
\usepackage{amsmath}
\usepackage{comment,xcolor}
\usepackage{hyperref}
\usepackage{gensymb}
\usepackage{xspace}
\usepackage{array}
\usepackage{cuted}
\usepackage{orchidlink}
\usepackage{tikz}
\usepackage{booktabs}  % For improved table formatting
\usepackage{subcaption}  % Include this in the preamble
\usepackage{afterpage}  % Include this in the preamble
\usepackage{colortbl}
\usepackage{rotating}   % For sidewaystable
\usepackage{xcolor}     % For row colors
\usepackage{colortbl}   % For cell coloring
\usepackage{booktabs}   % For cleaner tables (optional)
\usepackage{pdflscape}   % For proper landscape orientation
\usepackage{colortbl}    % For better table shading
\usepackage{booktabs}    % For cleaner tables
\usepackage{array}       % For column width adjustments
\usepackage{longtable}
\usepackage{booktabs}
\usepackage{multirow}
\usepackage{rotating}
\usepackage{adjustbox}

\definecolor{dg}{rgb}{0.0, 0.6, 0.1}

\title[Comparison of Galactic magnetic field models and synchrotron emission]{A comparison between Galactic magnetic field models and polarized synchrotron emission with C-BASS at 4.76\,GHz and S-PASS at 2.3\,GHz}

\author [V.Shaw et al]{Vasundhara~Shaw$^{1,6}$\thanks{E-mail: vasundhara.shaw@fiz-karlsruhe.de}\orcidlink{0000-0002-5824-7191},
    S.\,E.~Harper,$\!^1$,
    C.~Dickinson,$\!^{1}$\thanks{E-mail: clive.dickinson@manchester.ac.uk}
    J.\,P.~Leahy,$\!^1$\orcidlink{0000-0003-2514-9592}
    Gabriel~A.~Hoerning,$^{\!1}$\orcidlink{0000-0002-8677-6656}\newauthor
    R.~Cepeda-Arroita,$\!^{1,2}$
    Gilles Weymann-Despres,$\!^3$\orcidlink{0000-0002-9281-281X}
    Mike Peel,$\!^{4}$ \orcidlink{0000-0003-3412-2586}
    Angela C. Taylor,$\!^{3}$\orcidlink{0000-0002-3309-9081}\newauthor
    T. J. Pearson,$\!^{5}$\orcidlink{0000-0001-5213-6231}
    Jamie Leech,$\,^{3}$
    Michael Jones$\!^{3}$\\
$^1$Jodrell Bank Centre for Astrophysics, Alan Turing Building, Department of Physics \& Astronomy, School of Natural Sciences, \\ The University of Manchester,
Oxford Road, Manchester, M13 9PL, United Kingdom. \\ 
$^{2}$Instituto de Astrof\'{i}sica de Canarias, 38200 La Laguna, Tenerife, Canary Islands, Spain.\\
$^{3}$Department of Physics, University of Oxford, Denys Wilkinson Building, Keble Road, Oxford OX1 3RH, United Kingdom.\\
$^{4}$Imperial College London, Blackett Lab, Prince Consort Road, London SW7 2AZ, United Kingdom.\\
$^{5}$Owens Valley Radio Observatory, California Institute of Technology, Pasadena, CA 91125, USA.\\
$^{6}$FIZ Karlsruhe – Leibniz-Institute for Information Infrastructure, Franklinstr. 11, 10587 Berlin, Germany.
}

\begin{document}
\date{Accepted XXX. Received YYY; in original form ZZZ}

\pagerange{\pageref{firstpage}--\pageref{lastpage}} \pubyear{2026}

\maketitle

\begin{abstract}
We compare a set of contemporary Galactic magnetic field (GMF) models with polarized
synchrotron observations from the S-PASS and C-BASS radio surveys and combine them to create a reconstructed 4.76~GHz full sky map. Pixels that potentially have a large Faraday rotation are excluded while small ($< 80\degree$) Faraday corrections derived at the respective frequencies of the two surveys are applied to the rest of the map.  Using a template-fitting approach, we evaluate the ability of each model to reproduce the observed polarization amplitudes and polarization angles. We find that while most GMF models match the polarization angles reasonably well, they often fail to reproduce the morphology of the polarized intensity. We find that for most models there is a clear correlation between the data and models in polarization angles on large scales, but this does not hold true for polarized intensity. Our results show that a large portion of the polarized sky is shaped by local ``foreground'' features such as the North Polar Spur/Loop\,I and the Fan region. We conclude that incorporating such local structures is essential for accurately modelling the polarized synchrotron emission at microwave frequencies.
\end{abstract}


\section{Introduction}

In recent years, there has been a renewed focus on mapping diffuse Galactic emission, driven by two key motivations: the need to subtract foreground contamination in high-sensitivity Cosmic Microwave Background (CMB) experiments \citep[e.g.,][]{Dunkley2009a, Errard2016}, and the broader goal of advancing our understanding of the interstellar medium (ISM) and Galactic structure \citep[e.g.,][]{Carretti2013, Vidal2015, PIP_XII}. While numerous radio surveys have provided total intensity maps over a broad frequency range \citep[see e.g.,][]{deOliveira-Costa2008}, high-resolution, wide-area polarization data remain relatively scarce.
{Wilkinson Microwave Anisotropy Probe} (WMAP) and \textit{Planck} provide data at high frequencies, which helps to rule out Faraday rotation but are far less sensitive than more modern instruments such as { C-Band All-Sky Survey} (C-BASS)  \citep{WMAP_2004,WMAP_Page,Planck_NPIPE,Planck_X}.

Diffuse polarized {emission} from the Galaxy arises from several physical mechanisms, each tracing different components of the Galactic magnetic field (GMF). 
\textit{Synchrotron radiation}, produced by relativistic electrons spiralling around magnetic field lines, dominates at radio frequencies (typically below 100\,GHz) and probes the magnetic field component perpendicular to the line-of-sight ($B_{\perp}$). \textit{Dust polarization}, originating from thermal emission by non-spherical dust grains aligned with the magnetic field, also traces the plane-of-sky magnetic field orientation, particularly at far-infrared and sub-millimeter wavelengths. \textit{Starlight polarization}, observed at optical and near-infrared wavelengths, is caused by the differential extinction of background starlight by aligned dust grains and provides complementary constraints on the plane-of-sky magnetic field geometry \citep{Rybicki_book}. \textit{Faraday rotation}, the rotation of the plane of polarization of synchrotron emission as it propagates through the diffuse ionised medium, provides a probe of the line-of-sight component of the magnetic field ($B_{\parallel}$). This effect can also lead to a reduction in the observed degree of polarization, known as \textit{Faraday depolarization} \citep{Longair,Rybicki_book}.

Several recent and ongoing surveys have been dedicated to mapping these polarized signals across a wide range of frequencies and angular resolutions. Notable among them are the S-band Parkes All-Sky Survey (S-PASS; 2.3\,GHz) \citep{Carretti_2013,Carretti2019}, the C-Band All-Sky Survey (C-BASS; 4.76\,GHz) \citep{King:2009,Jones2018}, and the QUIJOTE experiment (10–40\,GHz), with published polarization measurements at 11--19\,GHz) \citep{Genova-Santos2015a, mfiwidesurvey}, which provide crucial measurements of synchrotron polarization. In addition, WMAP (23--94\,GHz) \citep{WMAP_2004} and \textit{Planck} (30--857\,GHz) \citep{Planck2015_XXV, Planck2018_IV} deliver all-sky coverage at higher frequencies, spanning both polarized synchrotron and thermal dust emission. Other high-resolution spectral surveys pursuing this goal at lower frequencies include the Global Magneto-Ionic Medium Survey (GMIMS; 300–1750\,MHz) \citep{Wolleben2009,Wolleben2019,Wolleben2021,sun2025,Ordog2025}, the polarization Sky Survey of the Universe’s Magnetism (POSSUM; 800–1088\,MHz) \citep{POSSSUM_pol}, and the Galactic-ALFA Continuum Transit Survey (GALFACTS 1225–1525\,MHz) \citep{Taylor2010}.

Complementary efforts targeting dust polarization at submillimeter wavelengths include the balloon-borne BLASTPol \citep{Fissel2016} and PILOT \citep{Bernard2016} experiments, as well as the JCMT BISTRO survey using SCUBA-2/POL-2 \citep{Ward-Thompson2017}, which achieve higher angular resolution in specific regions. At optical and near-infrared wavelengths, starlight polarization surveys such as GPIPS \citep{Clemens2012} and the ongoing PASIPHAE (Polar-Areas Stellar-Imaging in Polarization High-Accuracy Experiment) project \citep{pasiphae} provide independent constraints on the plane-of-sky magnetic field orientation. Together, these datasets provide critical insights into the polarized sky across multiple frequencies and tracers, forming the foundation for GMF modelling.

GMFs play a fundamental role in the ISM, influencing key processes such as turbulence, star formation, and cosmic ray propagation \citep{Federrath2012,Amato2017}. However, because of the absence of direct measurements, the GMF must be inferred indirectly by combining observations of synchrotron and dust polarization with models of the underlying electron and dust distributions. Modelling efforts aim to reconstruct both the large-scale coherent field and the small-scale turbulent components by integrating the emission along the line-of-sight to match observed data.

A number of GMF models have been developed using constraints from synchrotron intensity and polarization, Faraday rotation measures (RMs), and dust polarization \citep[e.g.][]{Sun_2008, Ruiz_Granados_2010, Jaffe_2010, PT11_2011, Jaffe_2011}, and even from populations of supernova remnant shells \citep{Mertsch2013}. However, our knowledge of the magnetic field in the Galactic halo remains considerably more limited than in the disc, mainly due to the scarcity of observational data at high Galactic latitudes and across wide frequencies \citep{Han1997antisymmetric,Han1999pulsar}. Nevertheless, several studies have found evidence for toroidal, axisymmetric field structures in the halo, which are anti-symmetric with respect to the Galactic plane, that is, the field reverses direction above and below the plane \citep{WMAP_Page, Sun_2008, Jaffe_2010,Ruiz_Granados_2010, PT11_2011, Han_2017,Han_2019}.

Further support for such halo field configurations comes from observations of external edge-on galaxies, many of which exhibit prominent X-shaped magnetic field structures \citep{Krause_2009, Beck_2009}. These observations have inspired analogous modelling efforts for the Milky Way. The widely used \textit{JF12} model \citep{JF12} incorporates both toroidal and X-shaped components and has served as a standard reference in the field. More recently, the \textit{UF23} suite of models \citep{Unger_2024} introduced eight distinct GMF configurations, incorporating greater structural complexity and improved fits to all-sky synchrotron data. These models also account for uncertainties in the distributions of thermal electrons and cosmic ray populations, offering a more flexible and comprehensive framework for interpreting polarized emission. In parallel, the \textit{KST24} model \citep{Korochkin_2025} marks an important advance by explicitly incorporating local structures such as the Local Bubble and the Fan region alongside the large-scale GMF \citep{West_2021,Dickey2022,Booth2025}. This model demonstrates an improved fit to WMAP polarization data and highlights the significance of accounting for nearby magnetic features that can substantially influence the observed sky signal.

These developments underscore the evolving landscape of GMF modelling and the crucial role of high-quality polarization data in constraining the field's structure particularly in the poorly understood halo region. In this work, we carry out a large-scale comparison of several state-of-the-art GMF models against the 4.76\,GHz C-BASS and 2.3\,GHz S-PASS data.

The structure of this paper is as follows: in Section~\ref{sec:data_discussion}, we introduce the C-BASS and S-PASS datasets and describe the masks used in the analysis. Section~\ref{sec:gmf_models} outlines the GMF models considered, detailing their individual components. In Section~\ref{sec:methodology}, we describe the computation of polarized synchrotron emission. Section~\ref{sec:results} presents the results of our template-fitting analysis. {Section~\ref{sec:strucfunc} presents the results of the structure function analysis, and in Section~\ref{sec:discussion}, we discuss the major results of the paper. Finally, in Section~\ref{sec:conclusion}, we outline the implications of our findings and directions for future work.}

\section{Combined full sky reconstructed synchrotron map}
\label{sec:data_discussion}

Currently no single full-sky polarized synchrotron map exists that provides both high sensitivity and minimal Faraday contamination. The only available full-sky datasets, WMAP and \textit{Planck}, are limited by their relatively low signal-to-noise ratio ($S/N$) for Galactic emission, whereas lower-frequency surveys, {in particular GMIMS \citep{Wolleben2009,sun2025,Ordog2025}}, which provides near all-sky polarization maps at 1.3–1.8\,GHz, are more strongly affected by Faraday depolarization. We combine the 4.76\,GHz C-BASS North and 2.3\,GHz S-PASS South polarized data to achieve the most complete and reliable reconstructed sky coverage currently possible at 4.76\,GHz.

\subsection{C-BASS North}
%{Introduction to the data}

\begin{figure}
    \centering
    \includegraphics[width = 1.0\linewidth]{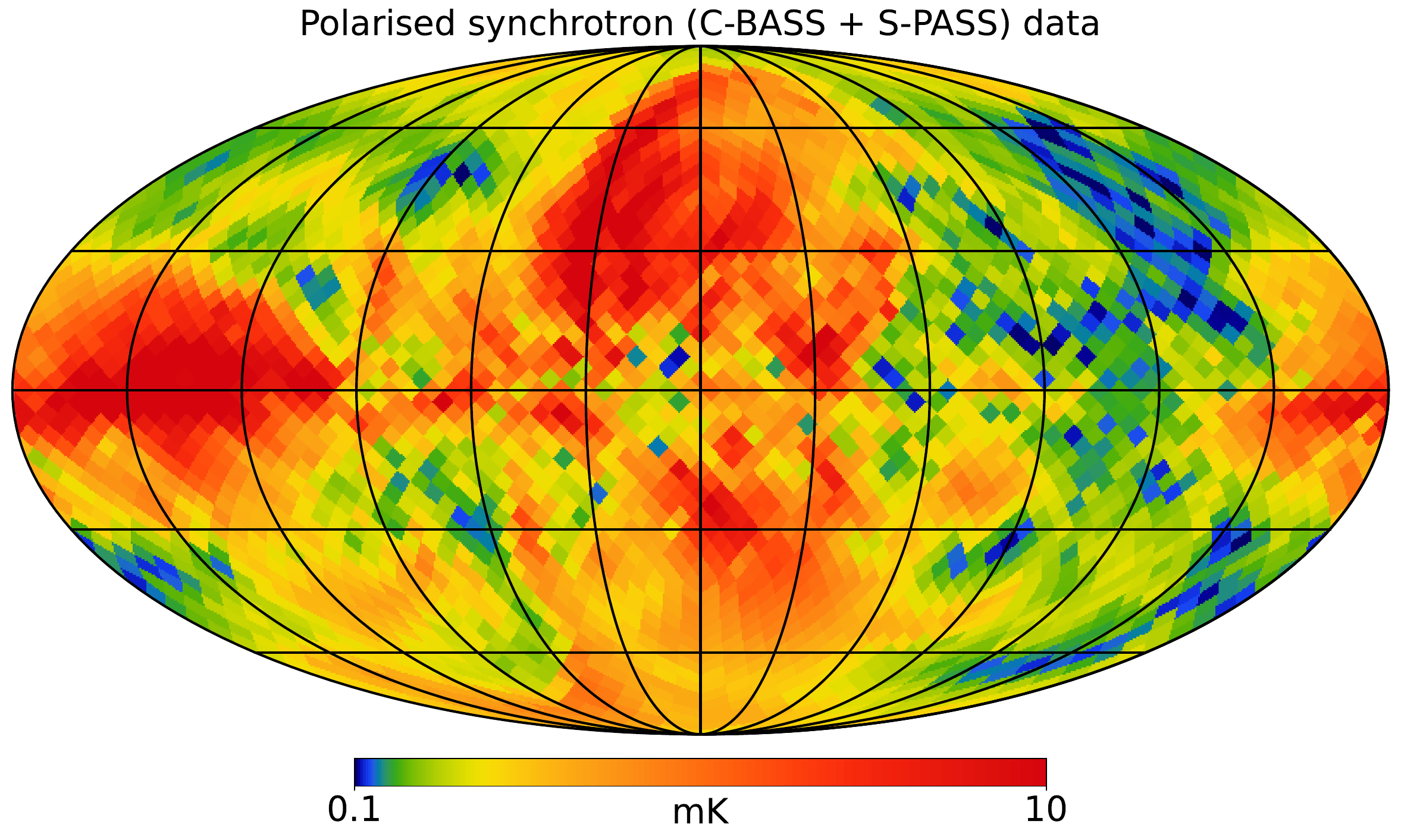}

    \caption{The combined C-BASS and S-PASS polarized synchrotron intensity map at 4.76\,GHz shown in Galactic coordinates and rendered at HEALPix resolution \( N_{\rm side} = 16 \) \citep{Gorski2005}.}
    \label{fig:franken_map}
\end{figure}

In this study we use the \textit{Data release 1} version of the C-Band All Sky Survey (C-BASS), an all-sky continuum survey measuring total intensity and linear polarization (Stokes $Q$ and $U$) at 4.76\,GHz (Taylor et al., in preparation). The C-BASS northern survey data were acquired using a 6.1\,m telescope at the Owens Valley Radio Observatory (OVRO) in California, USA, covering the sky north of declination $-15\fdg6$. C-BASS provides an angular resolution of 44\,arcmin (FWHM), comparable to that of the 408\,MHz Haslam map \citep{Haslam1982}, WMAP K-band \citep{WMAP_2004} and \textit{Planck} 30\,GHz LFI \citep[]{Planck2018_I}. As outlined by \citet{Jones2018}, the primary objectives of the survey are to support Cosmic Microwave Background (CMB) studies through improved foreground modelling, especially in polarization \citep{Jew2019}, and to investigate Galactic radio emission, which at this frequency is dominated by synchrotron and free-free radiation. The brightness temperature and polarization angle were calibrated using Tau A as a reference source \citep{Taylor2018}, Taylor et al. (in preparation), achieving accuracies better than 3\% in brightness temperature and within $1^{\circ}$ in polarization angle.
\subsection{S-PASS}

The second dataset employed in this work is the S-band polarization All Sky Survey (S-PASS), which provides measurements of the polarized radio emission from the southern hemisphere at a frequency of 2.3\,GHz \citep{Carretti2013, Carretti2019}. As the C-BASS southern data are not yet available, we compensate for its absence by scaling the S-PASS Stokes \textit{Q} and \textit{U} maps. S-PASS covers the sky at declination $\lesssim -1^\circ$ with an angular resolution of 8.9~arcmin, using observations from the 64-m Murriyang radio telescope at Parkes in New South Wales, Australia. The survey was designed to probe the magnetized interstellar medium at a frequency high enough to reduce strong depolarization at intermediate Galactic latitudes (still present at 1.4\,GHz), while remaining low enough to retain a high  $S/N$ at high latitudes for extragalactic and cosmological applications.

The final maps have an average sensitivity of 0.81~mK at beam-sized scales and show clear polarized signal across the southern sky, with typical emission values of 13\,mK in low-emission regions and $98.6\%$ of pixels exhibiting $S/N >3$ for {$N_{\rm{side}} = 1024$} \citep{Carretti2019}. The inner Galaxy hosts the strongest depolarization features, particularly near the Sagittarius arm. Due to the relatively low observing frequency, S-PASS measurements are significantly affected by Faraday rotation, which scales as $\nu^{-2}$. This effect is most pronounced at low Galactic latitudes ($|b| \lesssim 20^\circ$), where the magnetized plasma density is high, but becomes less influential at higher latitudes.

The S-PASS data products, including the calibrated Stokes \textit{Q} and \textit{U} maps, are publicly available and have been processed to mitigate systematic effects such as ground emission and instrumental leakage. These maps are a valuable resource for tracing the orientation and fluctuations of the GMF via polarization angle and Faraday depth studies.

\subsection{RM corrections, map combination and masking}
\label{sec:masks}

A carefully constructed masking scheme was applied to mitigate the influence of Faraday rotation and depolarization effects, which can significantly bias polarized synchrotron measurements. Since both C-BASS and S-PASS observations are affected by Faraday rotation, particularly the latter due to its lower observing frequency, rotation measure (RM) maps from Harper et al. (in preparation) were used to estimate and correct for the Faraday rotation angle in both C-BASS and S-PASS datasets independently. 
{The RM map was derived from a per-pixel linear fit between the C-BASS/S-PASS measured polarization angles with those measured by WMAP and Planck at 22.5 and 28.4~GHz.
We assumed that the observed polarization angle rotates as
\begin{equation} \label{RM_formula_1}
    \theta_{\rm{wl}} = \lambda^2\,RM + \theta_0,
\end{equation}
where $\theta_{\rm{wl}}$ is the polarization angle measured by the survey at wavelength $\lambda$ and $\theta_0$ is the intrinsic polarization angle. This form is consistent with both a foreground Faraday screen and the angle-rotation component of a Burn slab model with uniformly mixed emission and rotation \citep{Burn1966}, though the latter additionally predicts a reduction in polarized intensity. We performed the fit at 4 degree resolution (HEALPix $N_\mathrm{side} = 16$). Even at this resolution, some pixels in the WMAP/Planck data have very low signal-to-noise, so we imposed a prior on each pixel based on the RM map derived from extragalactic polarized sources \citep{Hutschenreuter2022}. We use the RM map derived from the fitted data rather than the \citet{Hutschenreuter2022} model alone because the diffuse Galactic polarized emission observed at radio frequencies is not necessarily located behind the full Faraday-rotating screen of the Galaxy. The data-driven RM map therefore provides a more appropriate description of the observed Faraday rotation environment at radio frequencies.

We note that this correction primarily serves to align the large-scale polarization 
angles with WMAP/Planck: the extrapolation from $\lambda \approx 1.3$\,cm to  $\lambda \approx 0$\,cm represents only $\sim$4\% of the $\lambda^2$ gap between C-BASS 
($\lambda \approx 6.3$\,cm) and WMAP/Planck ($\lambda \approx 1.3$\,cm). Consequently, 
the corrected position-angle errors are essentially the same as, and in fact marginally 
worse than, those of WMAP/Planck alone, with no advantage gained from C-BASS at the 
scale of $N_\mathrm{side} = 16$ pixels. This limitation applies only to polarization 
angles; the polarized intensity and fractional polarization, which form the basis of our 
GMF model comparisons, are unaffected.

}

Following this correction, a threshold was imposed: all pixels with \(\Theta_{\rm RM} > 80^\circ\), where $\Theta_{\rm RM}$ is the rotation angle,  were excluded from the analysis to ensure the fidelity of the recovered polarization angles. This threshold was applied separately to each map, and the resulting masks were combined to form an all-sky RM-based exclusion mask. We also re-ran the analysis with  stricter thresholds of \(\Theta_{\rm RM} > 50^\circ\) and \(\Theta_{\rm RM} > 20^\circ\)  and found no significant differences in the results. Additionally, we apply a mask for regions with poor  $S/N$:
\begin{equation}
P' = \sqrt{Q^2 + U^2}, \quad
P = \sqrt{P'^2 - \sigma_P^2},\quad
\mathrm{S/N} = \frac{P'}{\sigma_P},  
\end{equation}
{where $Q$ and $U$ are the maps from the combined C-BASS + S-PASS data, $P'$ is the measured polarized intensity whereas, $P$ is the noise de-biased polarized intensity and \(\sigma_P^2\) is its variance, following the \cite{Wardle1974} method. C-BASS has high $S/N$ across most of the sky; we mask any pixel with $S/N < 10$ 
to exclude the low-sensitivity region in the north-west.}

\begin{figure}
    \centering

    \includegraphics[width = 0.95\linewidth]{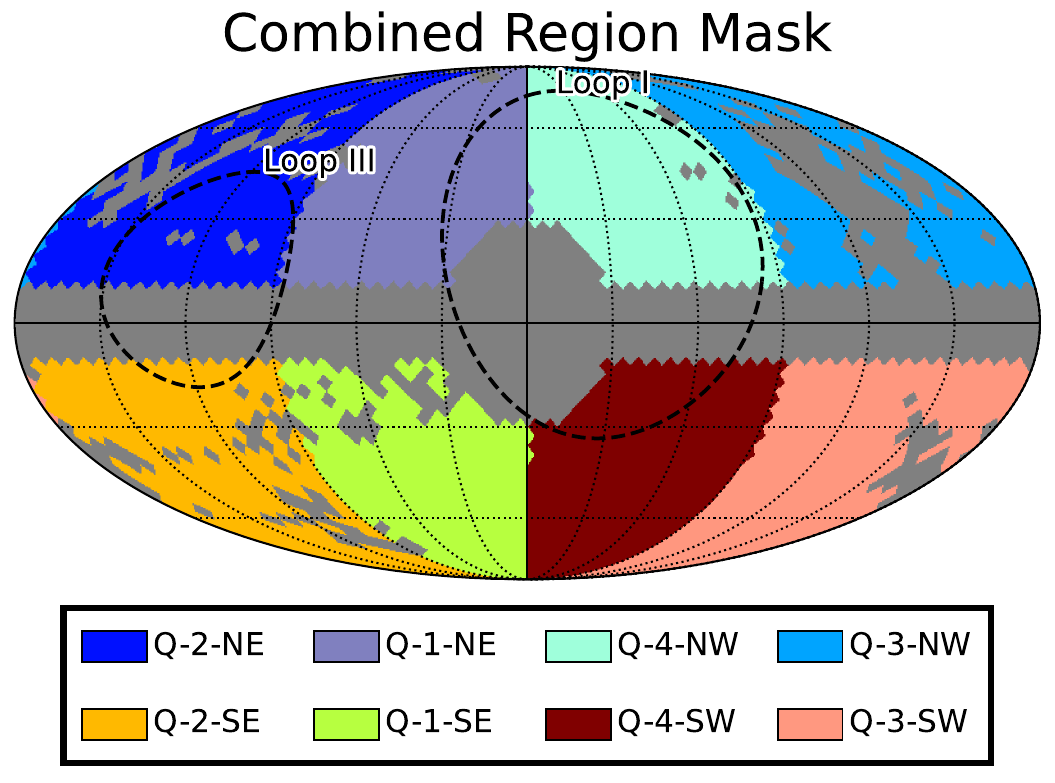}

    \caption{
    The \textit{Base map} and additional masked regions shown above in Galactic coordinates and rendered at HEALPix resolution \( N_{\rm side} = 16 \). The grey areas in the mask indicate excluded regions: a circular region of radius \( 30^\circ \)  centred on the Galactic Centre, where strong Faraday rotation leads to significant depolarization; a symmetric Galactic latitude cut of \( \pm10^\circ \) to remove bright disc emission; and high-latitude pixels with rotation measures angles exceeding \( 80^\circ \), which are also excluded due to Faraday contamination. The colour-coded regions denote specific subdivisions used for spatially resolved template fitting: \textit{Q-1-NE} through \textit{Q-4-SW}. Each quadrant (\textit{Q-NE} through \textit{Q-SW}) consists of two subregions (e.g., \textit{Q-1-NE} and \textit{Q-2-NE}). These quadrants are grouped to define broader sky masks: \textit{Q-NE} and \textit{Q-NW} form the \textit{North}, \textit{Q-SE} and \textit{Q-SW} form the \textit{South}, \textit{Q-NE} and \textit{Q-SE} define the \textit{East}, and \textit{Q-NW} and \textit{Q-SW} define the \textit{West}. The union of all these regions constitutes the \textit{Base mask}, which serves as the effective sky coverage, forming the \textit{Full sky} region, used consistently throughout all model comparisons.}
    \label{fig:region_masks}
\end{figure}

In addition to the RM-based masking, geometric exclusions were applied to remove regions of known complexity and strong depolarization. These include a circular region of radius $30^\circ$ centred on the Galactic Centre, as well as a $\pm10^\circ$ wide band around the Galactic plane. These regions are strongly affected by Faraday depolarization and are therefore unsuitable for direct comparison with model predictions. Together, these criteria define the \textit{Base mask}, which is uniformly applied to the combined C-BASS/S-PASS data as well as to all GMF model outputs in this work. The resulting sky coverage after masking is illustrated in Figure~\ref{fig:region_masks}.

To combine the C-BASS and S-PASS datasets, we extrapolate the S-PASS Stokes $Q$ and $U$ maps to 4.76\,GHz, assuming a uniform spectral index of $\beta = -3.1$ \citep{Fuskeland2014,Planck_XXV} and the scaling factor applied is $(4.76/2.303)^{\beta}$. We use the C-BASS dataset in the overlap region with S-PASS (declinations $-15^\circ$ to $-1^\circ$). {A comparison between the C-BASS and S-PASS dataset for the overlap region has been discussed in detail in Appendix~\ref{sec:appendixC}.} The extrapolated S-PASS maps and the C-BASS maps are then combined to produce full-sky Stokes \textit{Q} and \textit{U} maps at 4.76\,GHz. From these, a reconstructed polarized intensity map is computed, shown in Figure~\ref{fig:franken_map}. We use de-biased maps for this analysis and adopt the  $1^{\circ}$ resolution of C-BASS data. The S-PASS map was also smoothed with a Gaussian beam of $1^{\circ}$ before combining with C-BASS.

{Unlike Stokes $I$, the frequency dependence of polarized intensity can be affected by Faraday depolarization and line-of-sight mixing, which do not necessarily follow a simple power law. Our masking strategy removes regions of strong Faraday complexity, and we assess the residual depolarization using the standard Burn slab model \citep{Burn1966}. For typical intermediate- and high-latitude rotation measures $|\mathrm{RM}| \sim 10$--$30\,\mathrm{rad\,m^{-2}}$, the predicted reduction in fractional polarization at 4.76\,GHz is $\lesssim 1$--$2\%$, and at 2.3\,GHz is $\sim$5--16\%. These levels of depolarization are negligible at C-BASS frequencies and at worst modest in S-PASS, and would have no significant impact on our comparisons to model predictions. Turbulent magnetic fields can in principle produce depolarization with negligible Faraday rotation, but would require very high field strengths to cause significant depolarization if the turbulent scales are much smaller than the line-of-sight depth. If turbulent scales are instead comparable to or larger than the line-of-sight depth, such fields would be expected to produce more structure in both RM maps \citep[e.g.][]{Hutschenreuter2022, Thomson2023} and polarization angle than is observed at high Galactic latitudes (see Section~\ref{sec:strucfunc}). Under these conditions, a power-law scaling of Stokes $Q$ and $U$ provides a reasonable first-order approximation.}

We further subdivide the unmasked sky into several distinct sub-regions to allow a more detailed comparison. These include four longitudinal quadrants labelled \textit{Q-NE} through \textit{Q-SW}, along with a pair of corresponding sub-regions \textit{Q-1-NE, Q-2-NE, Q-3-NW, Q-4-NW, Q-1-SE, Q-2-SE} and \textit{Q-3-SW, Q-4-SW} that partition quadrant \textit{Q-NE--Q-SW} based on differing morphological behaviour observed in the data. These subdivisions are shown in Figure~\ref{fig:region_masks}. We  also define a \textit{North} mask combining regions \textit{Q-NE} and \textit{Q-NW}, a \textit{South} mask defined by the union of \textit{Q-SE} and \textit{Q-SW}, an \textit{East} mask combining \textit{Q-NE} and \textit{Q-SE} and a \textit{West} mask combining \textit{Q-NW} and \textit{Q-SW}. Regions \textit{Q-1-NE} and  \textit{Q-4-NW} combined include Loop I, the Fan region predominantly sits in regions \textit{Q-2-NE} and \textit{Q-2-SE}. For a more detailed discussion on Galactic foregrounds see Section~5 in Cepeda-Arroita et al. (in preparation).

The rationale behind this regional division is driven by the complex morphology and large-scale structures on the order of tens of degrees in the polarized synchrotron sky. {Additionally, these also align with the Galactic quadrants, which define cardinal directions along our line of sight into the Galaxy.} Prominent features such as the Fan region (approximately bounded in Galactic coordinates as $l \approx 120^\circ \ \text{to} \ 160^\circ, \quad b \approx -15^\circ \ \text{to} \ +20^\circ,$ centred near $l \sim 130^\circ, \ b \sim +5^\circ$) and the North Polar Spur (NPS) or Loop I running roughly from $(l,b) \approx (25^\circ, 20^\circ)$ up to $\approx (330^\circ, 75\text{--}78^\circ)$, may originate from relatively local ($\lesssim1$\,kpc) structures superimposed onto the large-scale GMF. By evaluating the model–data agreement within each of these regions, we aim to identify how existing GMF models perform in different regions of the sky at 4.76\,GHz.

%% B-field model details
\section{Magnetic field models}
\label{sec:gmf_models}

Various models have been proposed to describe the large-scale and turbulent components of the GMF, each constrained by different datasets such as Faraday rotation measures, synchrotron polarization, and cosmic ray arrival directions \citep{Sun_2008,Sun_2010,Jaffe_2010,Jaffe_2011,PT11_2011,JF12,Jansson_2012,Katia_2014,Han_2019,Shaw22,Unger_2024,Korochkin_2025}. In this work, we compare different magnetic field configurations drawn from current literature to simulate polarized synchrotron emission and evaluate their consistency with the 4.76\,GHz radio synchrotron data.
{An alternative route to the analytic GMF models discussed here is to use high-resolution MHD cosmological simulations to generate the magnetic field in spiral galaxies \citep[e.g.][]{Pakmor2018,Reissl2019,Reissl2023}. These simulations address the complexity of the projected fields in the disk plane, which we deliberately avoid. In particular, \citet{Reissl2023} demonstrate reasonable agreement between the all-sky power spectrum of their simulated RM pattern and observations of our Galaxy. While impressive, their approach de-emphasizes the actual geometry of the field, and the simulations act as a black box, whose details depend on numerous approximations for sub-grid physics, including star formation and ISM turbulence. Their all-sky power spectra are dominated by the high RMs close to the Galactic plane, whereas our analysis addresses lines of sight at high latitude, where the Galactic halo makes a significant contribution.}

{Below, we summarize the key features of each analytic GMF model used in our analysis. All the models considered in this work are listed in } Table~\ref{tab:gmf_models_summary}. Most of the models listed below comprise multiple components of the GMF with different geometries (e.g., toroidal or poloidal) that can span large scales (on the order of a few kiloparsecs) or small scales (on the order of tens to hundreds of parsecs). 
%We also provide a tabulated list of each model and model components used in Tables~\ref{tab:models_comparison}.

\begin{table*}
\centering
\caption{Summary of Galactic magnetic field (GMF) models used in this work, listing the shortened model names adopted throughout the paper, the magnetic field components included in each model, and the corresponding references.}
\label{tab:gmf_models_summary}
\begin{tabular}{|l|l|l|}
\hline
Model (short name) 
& Included magnetic field components 
& Reference \\ 
\hline

SVT22 str 
& Toroidal halo (coherent) 
& \citep{Shaw22} \\ \hline

SVT22 str tur 
& Toroidal halo + small-scale turbulence 
& \citep{Shaw22} \\ \hline

JF12 full str 
& Disc + toroidal halo + X-shaped poloidal field 
& \citep{JF12} \\ \hline

JF12 full str tur 
& Disc + toroidal halo + X-shaped poloidal field + turbulence 
& \citep{JF12} \\ \hline

JF12 tor X 
& Toroidal halo + X-shaped poloidal field 
& \citep{JF12} \\ \hline

JF12 tor disc 
& Toroidal halo + disc field 
& \citep{JF12} \\ \hline

JF12 X disc 
& X-shaped poloidal field + disc field 
& \citep{JF12} \\ \hline

UF23 base 
& Disc + toroidal halo + X-shaped poloidal field 
& \citep{Unger_2024} \\ \hline

UF23 expX 
& Disc + toroidal halo + modified X-shaped poloidal field 
& \citep{Unger_2024} \\ \hline

UF23 spur 
& Disc (local spur) + toroidal halo + X-shaped poloidal field 
& \citep{Unger_2024} \\ \hline

UF23 twistX 
& Twisted combined toroidal and poloidal halo field 
& \citep{Unger_2024} \\ \hline

UF23 nebCor 
& Disc + toroidal halo + X-shaped poloidal field (electron--field anti-correlation) 
& \citep{Unger_2024} \\ \hline

UF23 neCL 
& Disc + toroidal halo + X-shaped poloidal field (alternative $n_e$ model) 
& \citep{Unger_2024} \\ \hline

UF23 cre10 
& Disc + toroidal halo + X-shaped poloidal field (extended halo) 
& \citep{Unger_2024} \\ \hline

UF23 synCG 
& Disc + toroidal halo + X-shaped poloidal field (alternative synchrotron model) 
& \citep{Unger_2024} \\ \hline

KST24 full 
& Local Bubble + disc + toroidal halo + X-shaped poloidal field 
& \citep{Korochkin_2025} \\ \hline

KST24 Galaxy 
& Disc + toroidal halo + X-shaped poloidal field 
& \citep{Korochkin_2025} \\ \hline

XH19 
& Disc (spiral with reversals) + bi-toroidal halo 
& \citep{Han_2019} \\ \hline

LogSpiral 
& Logarithmic spiral disc geometry (polarization angles only) 
& \citep{Page2007} \\ \hline

\end{tabular}
\end{table*}

% {define what it means of large scale}

\subsection{\textit{SVT22 model for Galactic halo}}

The simplest model included in our analysis is the \citep{Shaw22} model hereafter \textit{SVT22}, which focuses on describing the magnetic field structure in the Galactic halo. It consists of two components: \textit{1)} a large-scale, axisymmetric toroidal field with an exponential radial cut-off at 5\,kpc and an azimuthal cut-off at 6\,kpc, and a magnetic field strength of approximately $4\,\mu$G (\textit{SVT22 str}); and \textit{2)} a small-scale turbulent component with a Kolmogorov power spectrum, coherence length of 150\,pc, and strength of approximately $6\,\mu$G. The structured and turbulent field models together form \textit{SVT22 str tur} model. {Parameters for both versions of this model were constrained by fitting to the \textit{Planck} 30\,GHz polarized synchrotron data.} The model does not include a disc component and assumes a uniform cosmic-ray electron density in the halo region based on the WMAP model for electron distribution \citep{WMAP_Page}.

\subsection{\textit{Jansson and Farrar model 2012} (JF12 model)}

The \citep{JF12,Jansson_2012} model hereafter \textit{JF12} provides a comprehensive, data-constrained description of the GMF, based on a combination of Faraday rotation measures and synchrotron emission observations. The model divides the GMF into three large-scale, structured components: 
\textit{1)} a toroidal halo field, \textit{2)} an X-shaped poloidal field, and \textit{3)} a disc field.

\begin{enumerate}
   \item \textit{Toroidal halo.} This component consists of purely azimuthal fields in the northern and southern hemispheres, with the fields extending to about 4\,kpc above and below the Galactic plane. The radial extent in the north is roughly half of that in the south. The typical field strength is $\approx 1.4\,\mu$G in both hemispheres.
   
   \item \textit{X-field.} Motivated by edge-on spiral galaxy observations \citep{Beck_2009,Krause_2009}, the X-field is a poloidal, axisymmetric field extending out of the Galactic plane at an angle. It provides a vertical magnetic field component, particularly important at high latitudes.
   
   \item \textit{Disc field.} The disc field is spiral in structure and was informed by earlier work from \citet{Brown_2003,Brown_2007}, based on the NE2001 thermal electron density model \citep{Cordes_2002}.
\end{enumerate}

In addition to the coherent field components, the \textit{JF12} model includes small-scale turbulent fields with varying coherence lengths, as well as striated random components. The best-fit parameters were obtained by fitting to synchrotron total and polarized intensity from WMAP, as well as over 40,000 extragalactic rotation measures from the NVSS RM catalogue \citep{Taylor_2009}. Synchrotron predictions were computed using the GALPROP cosmic ray propagation code \citep{Orlando_2011,Orlando2013}.

In our analysis, we use multiple variants of the \textit{JF12} model: the fully structured version (\textit{JF12 full str}), the structured plus turbulent version (\textit{JF12 full str tur}), and partial combinations of the three main components.

\subsection{\textit{Unger and Farrar model 2023} (UF23 model)}
The  \citep{Unger_2024} (\textit{UF23}) form a group of 8 GMF models with divergence-free magnetic fields. 
The core model, referred to as the \textit{UF23 Base}, consists of a spiral disc field, a toroidal halo field, and a poloidal X-field. The model is defined within a cosmic-ray electron diffusion region that extends $\pm 6$\,kpc from the Galactic plane. It uses the YMW16 thermal electron density model \citep{Yao_2017} and a cosmic-ray electron profile given by the \textit{Dragon} code \citep{Evoli_2017}, which solves for the steady-state propagation of cosmic rays in the Galaxy, enabling a self-consistent calculation of synchrotron emissivity across the model volume. The magnetic field components are parametrized using simple functional forms, and their amplitudes and scales are tuned to best fit the 30\,GHz \textit{Planck} polarized synchrotron emission maps \citep{Planck2018_IV}.

To account for striated turbulent magnetic fields which enhance polarized emission without contributing to rotation measures, the \textit{UF23} models introduce a multiplicative \textit{striation factor}, denoted by $1 + {\beta'}$. Here, ${\beta'}$ quantifies the contribution of striated random fields aligned with the regular magnetic field. This effectively boosts the predicted synchrotron intensity and polarization to match the \textit{Planck} data without altering the overall morphology of the large-scale field (see Section \ref{sec:discussion}). In total, eight model variants are provided in the UF23 suite including \textit{UF23 Base}, each designed to test different physical assumptions:
\begin{enumerate}

    \item \textit{UF23 expX}: Instead of the standard radial cut-off, this model applies an exponential dependence to the mid-plane vertical component of the poloidal X-field.
    \item \textit{UF23 spur}: The spiral structure of the disc field implied by the Fourier spiral is reduced to a single local spur as seen also in the cosmological simulation of \citep{{Pakmor_2014}}, corresponding to the Orion arm.
    \item \textit{UF23 twistX}: The toroidal and poloidal halo fields are combined into a unified model, producing a twisted X-field structure.
    \item \textit{UF23 nebCor:} Introduces an anti-correlation between the thermal electron density and magnetic field strength, motivated by observations in galaxy clusters and interstellar regions.
    \item \textit{UF23 neCL:} Uses the NE2001 thermal electron density model \citep{Cordes_2002} instead of YMW16 \citep{Yao_2017}, which changes the morphology of the disc field due to different assumptions about the Galactic spiral arms.
    \item \textit{UF23 cre10:}  Increases the height of the cosmic-ray electron diffusion region from 6\,kpc to 10\,kpc, allowing for more extended emission from the halo.
    \item \textit{UF23 synCG}: The default synchrotron emission model is substituted with an estimate derived from the COSMOGLOBE analysis \citep{Watts2023,Watts2024}.
\end{enumerate}

\subsection{\textit{Korochkin, Semikoz and Tinyakov  model 2024} (KST24 model)}

The {\textit{KST24}} GMF model \citep{Korochkin_2025}  attempts, for the first time, to model the Local Bubble along with the large-scale structures of the Galaxy. The model is broadly divided into two components:
\begin{enumerate}
    \item \textit{KST24 Local Bubble:}  the authors approximate the Local Bubble to a spherical empty shell that sits in the disc of the Galaxy. The field strength is determined by the compression of magnetic field lines from an initially uniform field due to the expansion of a spherically symmetric bubble. The field inside the shell is zero and outside the shell remains unchanged, whereas the field in the wall of the shell is tangential to its surface and distributed uniformly through the wall in the radial direction.
    \item \textit{KST24 Galaxy:} this Galaxy model comprises: \textit{(1)} a thick disc that traces the Milky Way’s spiral arms, \textit{(2)} a toroidal halo magnetic field, and \textit{(3)} a symmetric X-shaped magnetic field with topological similarities to Model C in \citep{Katia_2014}. To reduce the number of free parameters, the authors assume the X-field lines are symmetric with respect to the Galactic plane, axially symmetric, and maintain a constant inclination angle $\theta$ relative to the Galactic axis.
\end{enumerate}
The two components combine to form the \textit{KST24 full} model, which we also use in our analysis. This model does not include any turbulent components.

\vspace{-0.58cm}

\subsection{\textit{Xu and Han model 2019 (XH19)}}
The model by \cite{Han_2019} (\textit{XH19}) focuses on the large-scale GMF structure in the solar vicinity and the Galactic halo. It is constructed using rotation measure data from pulsars and extragalactic radio sources. The disc field is modelled as a logarithmic spiral with multiple reversals, consistent with previous studies suggesting field direction changes between spiral arms and inter-arm regions. The authors find that a spiral configuration with at least three reversals best fits the RM data in the inner Galaxy.

The authors propose a model with a purely bi-toroidal halo field above and below the Galactic plane, with opposite directions in the northern and southern hemispheres \citep{Han1997antisymmetric,Han1999pulsar}. The field strength peaks at a few $\mu G$ and decays exponentially with Galactocentric radius and vertical distance from the plane. Unlike models that incorporate X-shaped or poloidal components, this model emphasizes a simpler, azimuthally dominant configuration that best fits the available RM observations in the nearby Galactic environment.

This relatively simple axisymmetric configuration provides a good fit to local RM observations and complements more complex models by offering insight into the magnetic field structure in the solar neighbourhood and lower halo. Note that since the publication of \cite{Han_2019} significant progress has been made in improving the (extragalactic) RM data \citep{POSSSUM_pol}.

\subsubsection{LogSpiral}
The Logarithmic Spiral Arm (or \textit{LogSpiral}) model \citep{Page2007} consists of a smooth spiral pattern in cylindrical coordinates, with the field geometry following a logarithmic spiral form. It includes a vertical tilt that increases with height above the Galactic plane. The model uses a few key parameters (like pitch angle and scale height) to match observed polarization angle data. This is a geometric model designed to reproduce the polarization-angle sky map; it requires arbitrary values of the magnetic field in order to produce any synchrotron emission. The authors of the paper provide no magnetic field strengths in their work. This model is mainly useful to compare polarization angles and we do not compare polarized synchrotron intensity from this model.

\section{Calculation of the polarized synchrotron skymaps from the magnetic field model}
\label{sec:methodology}
\subsection{Non-thermal electron distribution}
\label{sec:EdNdE_disc}

To generate synthetic synchrotron maps, both a model for the non-thermal or relativistic cosmic ray electron distribution and a magnetic field model are necessary. The \textit{JF12} model considered two different approaches for the non-thermal electron distribution: the WMAP analytical expression and a simulated distribution from GALPROP \citep{Strong2007}, ultimately adopting the latter. These models differ significantly, as the WMAP model \citep{WMAP_Page} is a semi-analytical formula, while the GALPROP model is derived theoretically by solving the diffusive transport equation under the assumption of a specific spatial source distribution with an absorptive halo boundary. Given the limited knowledge of the non-thermal electron distribution in the Galaxy, we opt for the simpler WMAP analytical model to minimise additional complexities for most of the analysis. The adopted WMAP non-thermal electron  density distribution as a function of galactocentric radius $r$ and distance from the disc plane $z$ is expressed as:

\begin{equation}\label{Eq_WMAP_EdNdE}
    \frac{\mathrm{d}n_e}{\mathrm{d}\log E_\mathrm{e}} = C_\mathrm{norm} \left(\frac{E_e}{E_{\rm 10\,GeV}}\right)^{-p+1} e^{-r/R_{\mathrm{el}}}\, \mathrm{sech}^2\left(\frac{z}{Z_{\mathrm{el}}}\right), 
\end{equation}
where $\frac{\mathrm{d}n_e}{\mathrm{d}\log E_\mathrm{e}}$ represents the differential electron spectrum in logarithmic energy bins measured in ${\rm cm}^{-3}$ {with $n_e$ and $E_e$ being the electron energy}. The spectral index of the electron spectrum is set to $p = 3$. The parameter $C_\mathrm{norm} \approx 1 \times 10^{-13}~\mathrm{cm^{-3}}$ similar to that adopted by \citep{JF12,Strong2007} defines the electron density at an energy of 10~GeV and is fixed, while $R_{\mathrm{el}}$ and $Z_{\mathrm{el}}$ set the radial and vertical spatial scale lengths at 5\,kpc and 6\,kpc respectively based on the estimates from the best fit values in \citet{Shaw22}.\\

For comparison we compared polarized intensity (PI) maps for the \textit{KST24} and \textit{UF23} models calculated from a different electron distribution code called \textit{Dragon} \citep{Dragon_2013,Dragon_2017}. The PI maps for both these models were provided to us via private communication by the respective authors \citep{Unger_2024,Korochkin_2025}. For clarity, PI maps computed  using \textit{Dragon} are labelled as (for example) \textit{KST24 full Dragon} and the PI map calculated using the WMAP electron distribution is named a \textit{KST24 full WMAP}. The same naming convention is also adopted for the case with \textit{UF23} models. We quantify the effect of different electron distributions in Section~\ref{sec:diff_EdNdE_disc}.

\subsection{Synchrotron emission}

Synchrotron radiation arises from relativistic charged particles gyrating around magnetic field lines sensitive to $B_{\perp}$, the component of the magnetic field perpendicular to the observer’s line-of-sight, and is intrinsically polarized \citep{Westfold,Rybicki_book}. The polarized emissivity can be described as an ellipse, with the major axis corresponding to the perpendicular component ($J_{\perp}$) and the minor axis to the parallel component ($J_{\parallel}$), see Appendix~C in \citet{Shaw22} for details.

The two components, $J_{\perp}$ and $J_{\parallel}$, describe the emission spectrum for a given peak photon energy $E_{\gamma}^{\mathrm{peak}}$. 
Expressions for these two components, produced by electrons between the energy ranges of $E_e^{\mathrm{min}} = 1$~GeV and $E_e^{\mathrm{max}} = 100$~GeV,  with pitch angle $B_{\perp}/B$ for each step along the line-of-sight ($l$), are:
% provided below in Eqs.~\ref{Jperp} and \ref{Jpara}, (Redundant - Paddy)
\begin{equation}
 J_{\perp}^l = \frac{1}{\tau}  \int_{\log E_e^{\mathrm{min}}}^{\log E_e^{\mathrm{max}}} \  \frac{\mathrm{d}n_e}{\mathrm{d}\log E_\mathrm{e}} \mathrm{d}\log E_\mathrm{e}\  \left[F\left(\frac{E_{\gamma}}{E_{\gamma}^{\mathrm{peak}}}\right) + G\left(\frac{E_{\gamma}}{E_{\gamma}^{\mathrm{peak}}}\right)\right] \\ 
 \label{Jperp}
\end{equation}

\noindent and

\begin{equation}
J_{\parallel}^l = \frac{1}{\tau} \int_{\log E_e^{\mathrm{min}}}^{\log E_e^{\mathrm{max}}} \ \frac{\mathrm{d}n_e}{\mathrm{d}\log E_\mathrm{e}} \mathrm{d}\log E_\mathrm{e}\  \left[F\left(\frac{E_{\gamma}}{E_{\gamma}^{\mathrm{peak}}}\right) - G\left(\frac{E_{\gamma}}{E_{\gamma}^{\mathrm{peak}}}\right)\right] 
\label{Jpara}
\end{equation}

where

\begin{align}
\tau^{-1} &= \frac{\sqrt{3} \alpha}{4\pi}\frac{B_{\perp}}{B_{\mathrm{crit}}}\frac{m_{e}c^{2}}{\hbar},
 & 
E_{\gamma}^{\mathrm{peak}} &= \frac{3}{2} \frac{E_\mathrm{e}^2}{m_{e}c^2} \frac{B_{\perp}}{B_{\mathrm{crit}}} ,
\nonumber
\end{align}
% E_{\gamma}^{\mathrm{peak}} &= \frac{3}{2}\Gamma_{e}^2 \frac{B_{\perp}}{B_{\mathrm{crit}}} m_{e} c^2,

and

\begin{align}
F(x) &= x \int_x^\infty K_{5/3}(x') dx', &
G(x) &= x K_{2/3}(x).
\nonumber
\end{align}
with $K_{5/3}$ and $K_{2/3}$ being the modified Bessel functions describing the shape of the photon spectrum. 

These expressions incorporate the critical magnetic field strength, $B_{\mathrm{crit}} = \frac{m_e^2c^3}{e\hbar} = 4.414 \times 10^{13}$~G, where $m_e c^{2} = 0.511$~MeV represents the electron rest energy, $h = 4.136 \times 10^{-15}$~eV~s is {Planck}'s constant with $\hbar = \frac{h}{2 \pi}$,  and $\alpha \approx \frac{1}{137.04}$ is the fine-structure constant.

The adopted conventions are outlined as follows. The parallel polarization component (${J_{\parallel}}$) is aligned with $\vec{B}_{\perp}$, whereas the perpendicular component (${J_{\perp}}$) is orthogonal to $\vec{B}_{\perp}$. The Stokes parameters at each line-of-sight point are defined in terms of the intrinsic polarization angle $\Psi^l_{\rm in}$, the angle between the line-of-sight perpendicular magnetic field $B_{\perp}$ and Galactic North. We adopt the IAU convention for calculating the angles, which is also adopted by C-BASS.

At each line-of-sight step, the components ${J_{\perp}^l}$ and ${J_{\parallel}^l}$ contribute to the Stokes parameters $Q$ and $U$ via:

\begin{eqnarray}
Q_{\rm in}^{\rm tot} = \frac{1}{4\pi} {\int_0^L \mathrm{d}l \ ({J_{\perp}^l} - J_{\parallel}^l) \ {\cos}(2\Psi^l_{\rm in}) }, \\
U_{\rm in}^{\rm tot} =\frac{1}{4\pi} {\int_0^L \mathrm{d}l \ ({J_{\perp}^l} - J_{\parallel}^l) \ {\sin}(2\Psi^l_{\rm in})}.
\end{eqnarray}

The polarized intensity $P_{\rm I}$ is then defined as:

\begin{equation} \label{eq_I_pol}
P_{\rm I} = \sqrt{(Q_{\rm in}^{\rm tot})^2+(U_{\rm in}^{\rm tot})^2}.
\end{equation}

Finally, the intrinsic polarization angle $\Psi_{\rm in}$ is obtained as:

\begin{equation}
\tan(2\Psi_{\rm in}) = \frac{U_{\rm in}^{\rm tot}}{Q_{\rm in}^{\rm tot}}.
\end{equation}

\begin{figure*}
    \centering
 
    \includegraphics[width=1.0\linewidth]{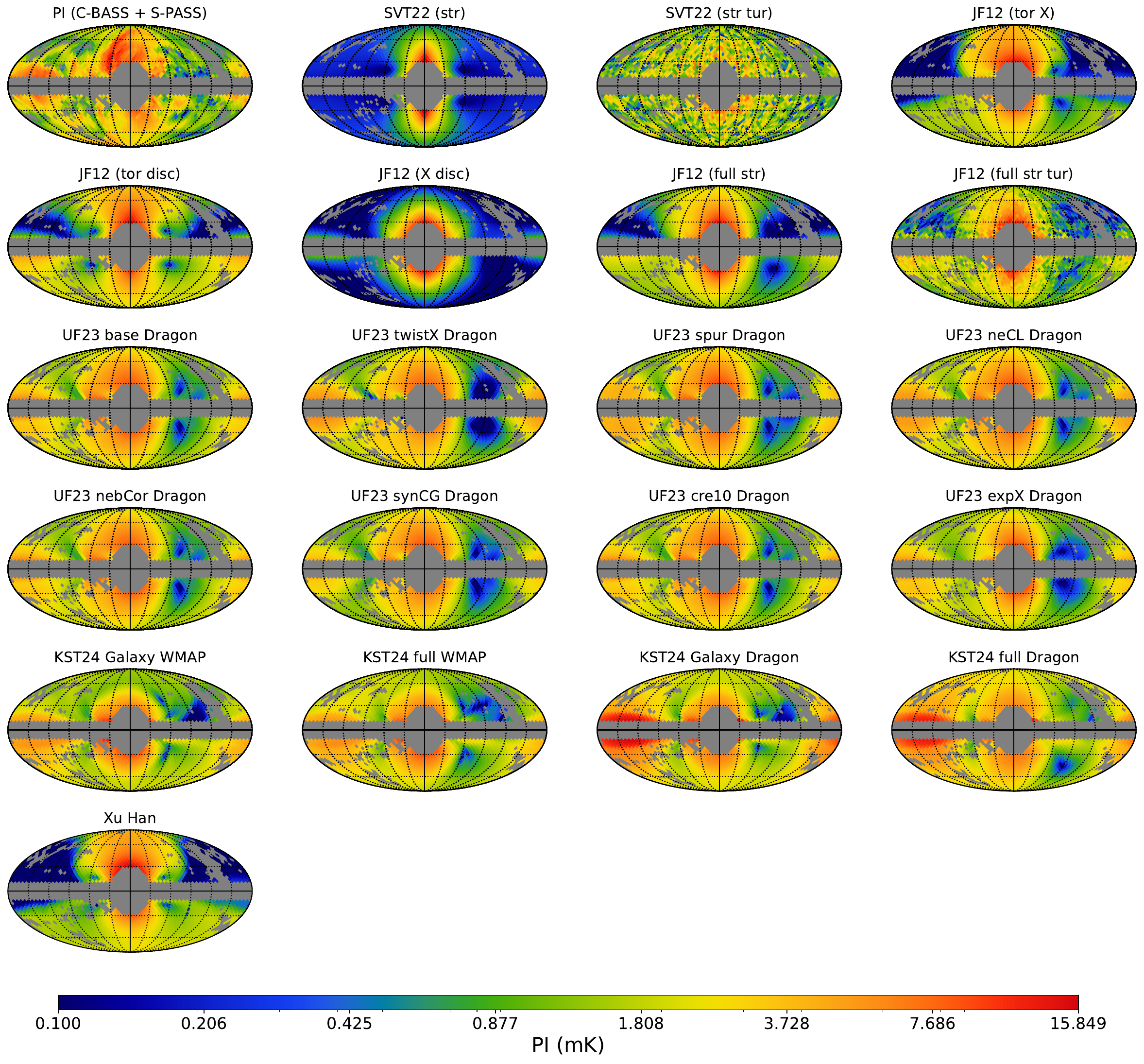}
    \caption{Polarized intensity (PI) synchrotron skymaps at 4.76\,GHz, presented in Galactic coordinates at HEALPix resolution \(N_{\rm side} = 16\). The skymap in the top-left panel shows the combined C-BASS/S-PASS observational data, followed by synthetic PI maps generated from each GMF model. All maps have been processed using the same \textit{Base mask} and have been rescaled by their respective fitted amplitudes obtained from fitting the full sky outside the \textit{Base mask}, allowing for direct comparison with the observed synchrotron intensity. A continuous logarithmic colour scale is used across all maps to enable consistent visual comparison. }
    \label{fig:PI_model_maps}
\end{figure*}

We calculate all the polarized synchrotron skymaps for $N_{\rm{side}}$ = 16 ($3.7^{\circ}$ pixel size) in Figure~\ref{fig:PI_model_maps} ($P_{\rm I}$) with the RM mask shown in Figure~\ref{fig:region_masks} for all the GMF models with the WMAP electron distribution model.

\begin{figure*}
    \centering

    \includegraphics[width=1.0\linewidth]{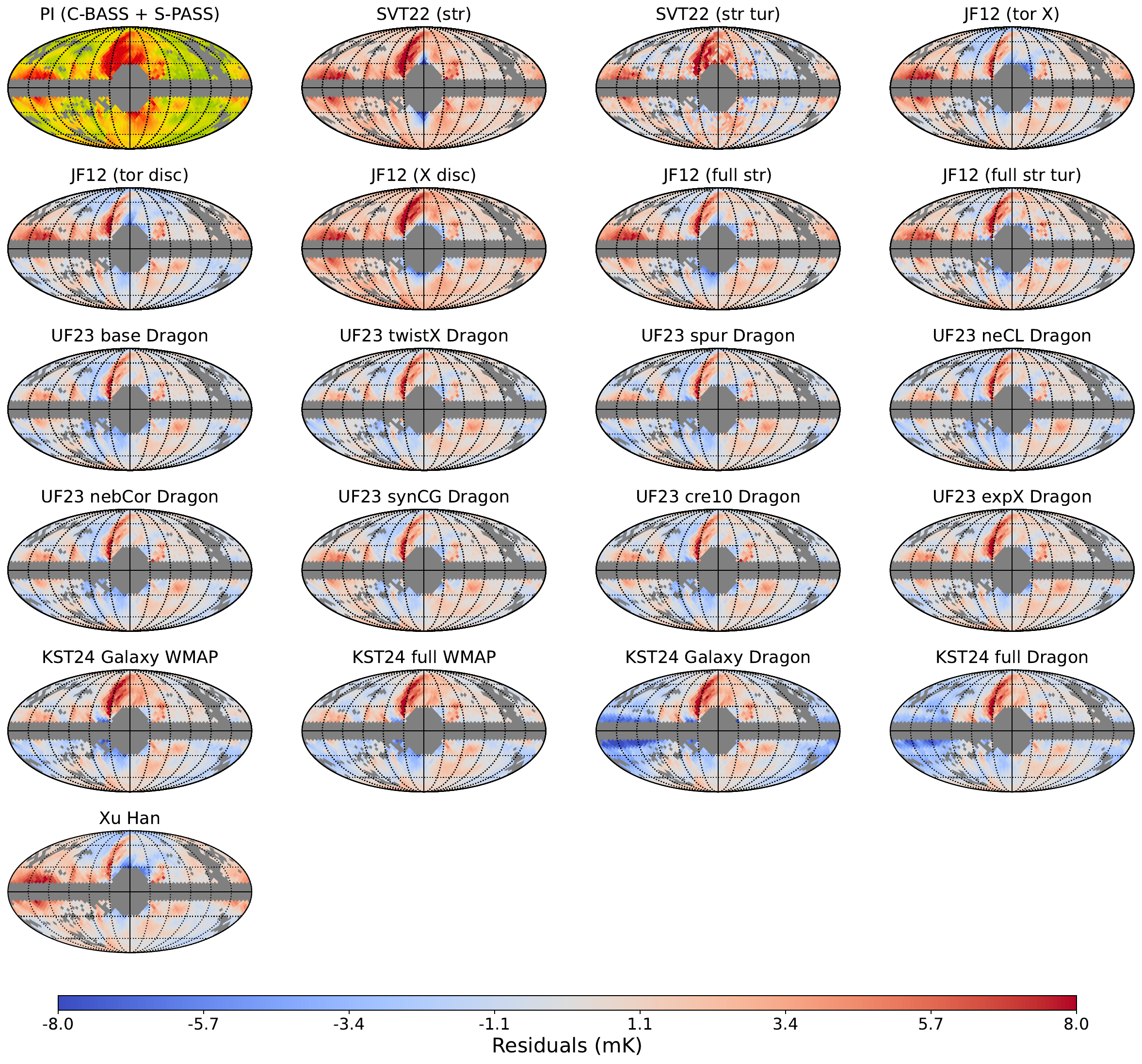}

    \caption{Residual polarized intensity maps at 4.76\,GHz, shown in Galactic coordinates at HEALPix resolution \(N_{\rm side} = 16\) with the \textit{Base mask} applied (masked regions appear in grey). For reference, the C-BASS data in the same units but with a different colour map are displayed in the top-left corner. The residuals are obtained by subtracting the modelled synchrotron intensity from the combined C-BASS/S-PASS observations. All maps use a consistent diverging colour scale: dark red regions indicate where the data are brighter than the model predictions, while blue regions mark areas where the model overestimates the observed intensity. Most residuals highlight the prominent NPS and the Fan region in red, as these are not incorporated in any of the models, except for the \textit{KST24} models which include the Fan region but not the NPS. }
    \label{fig:diff_PI_maps}
\end{figure*}

\begin{figure*}
    \centering
    \includegraphics[width=1.0\linewidth]{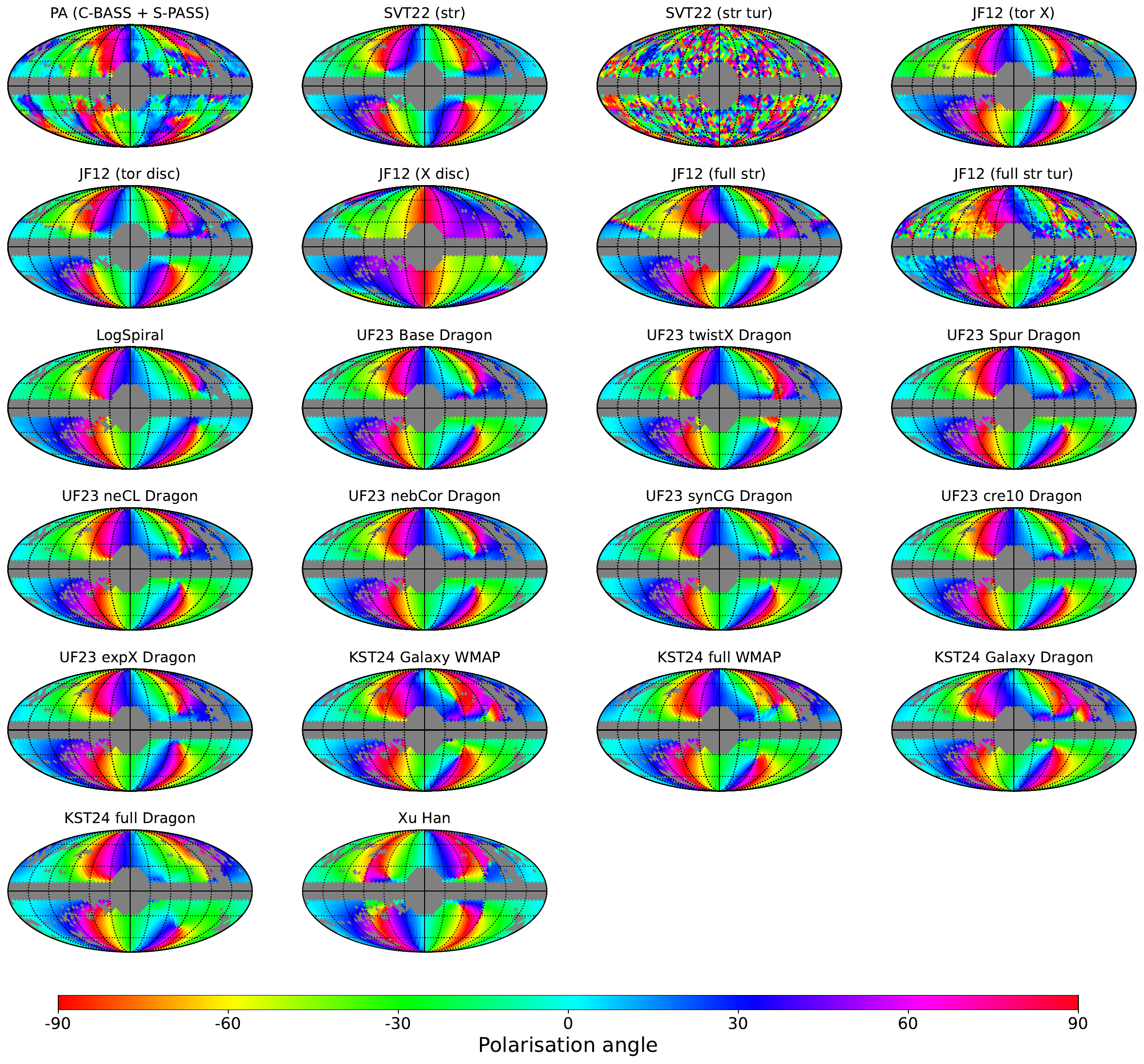}
    \caption{Polarization angles (PA) skymaps at 4.76\,GHz, presented in Galactic coordinates at HEALPix resolution \(N_{\rm side} = 16\). The skymap in the top-left panel shows the combined C-BASS/S-PASS observational data, followed by synthetic PA maps generated from each GMF model. All maps have been processed using the same \textit{Base mask} and a cyclic colour scale is used across all maps to enable consistent visual comparison of the angles. We show the \textit{LogSpiral} model only for this case where it is the most relevant.  }
    \label{fig:pol_angles}
\end{figure*}

\section{Results - Visual Comparisons of GMF Models with Synchrotron Maps}
\label{sec:results}

\begin{figure*}

    \includegraphics[width = 1.0\linewidth]{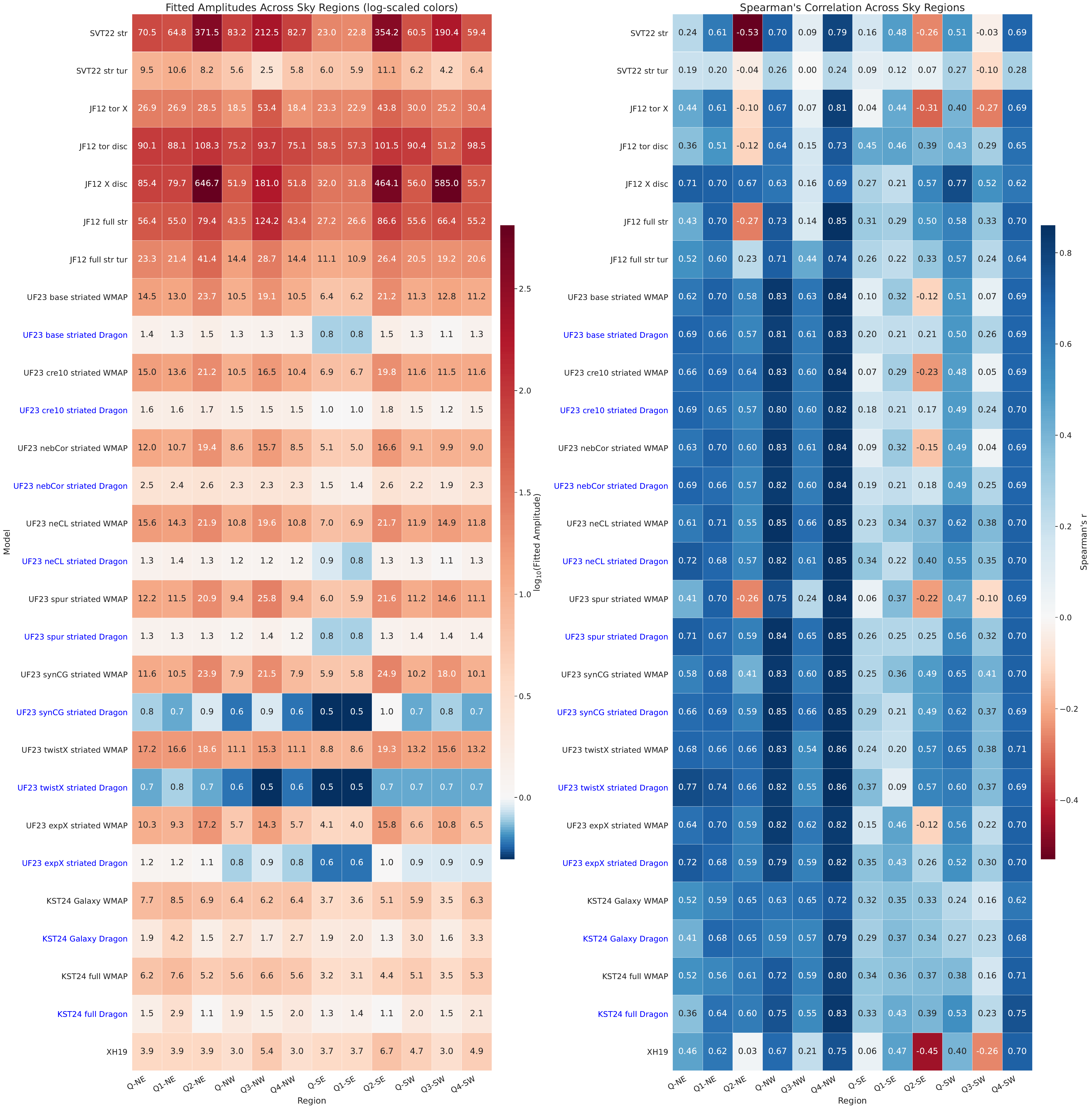}
\caption{{Heatmaps showing the fitted amplitude (left) and Spearman's rank correlation coefficient (right) for each GMF model across selected small sky regions, evaluated at HEALPix resolution \(N_{\mathrm{side}} = 16\). The fitted amplitude represents the scaling factor obtained from the linear template fit required to match the modelled polarized intensity to the observed C-BASS/S-PASS data at 4.76\,GHz. The Spearman's correlation coefficient quantifies the monotonic morphological agreement between model and data, independent of absolute amplitude. For the fitted amplitudes, the colour scale is logarithmic to accommodate the large dynamic range, while the annotated values are linear; red indicates lower amplitudes (better agreement in brightness) and blue indicates higher amplitudes (greater rescaling required). For the Spearman's correlation coefficients, the colour scale is linear; blue indicates higher correlation (better morphological agreement) and red indicates lower or negative correlation (poorer agreement). Both metrics are computed regionally using the masks defined in Figure~\ref{fig:region_masks}. All skymaps are calculated using the \textit{WMAP} electron distribution model except \textit{KST24 full/Galaxy Dragon}. We obtained the Stokes \textit{Q} and \textit{U} maps for the \textit{DRAGON} model from the authors of \citet{Unger_2024} via private communication.}}

    \label{tab:amp_Spearman_final_8Quad}
\end{figure*}

\begin{figure*}

\includegraphics[width = 1.0\linewidth]{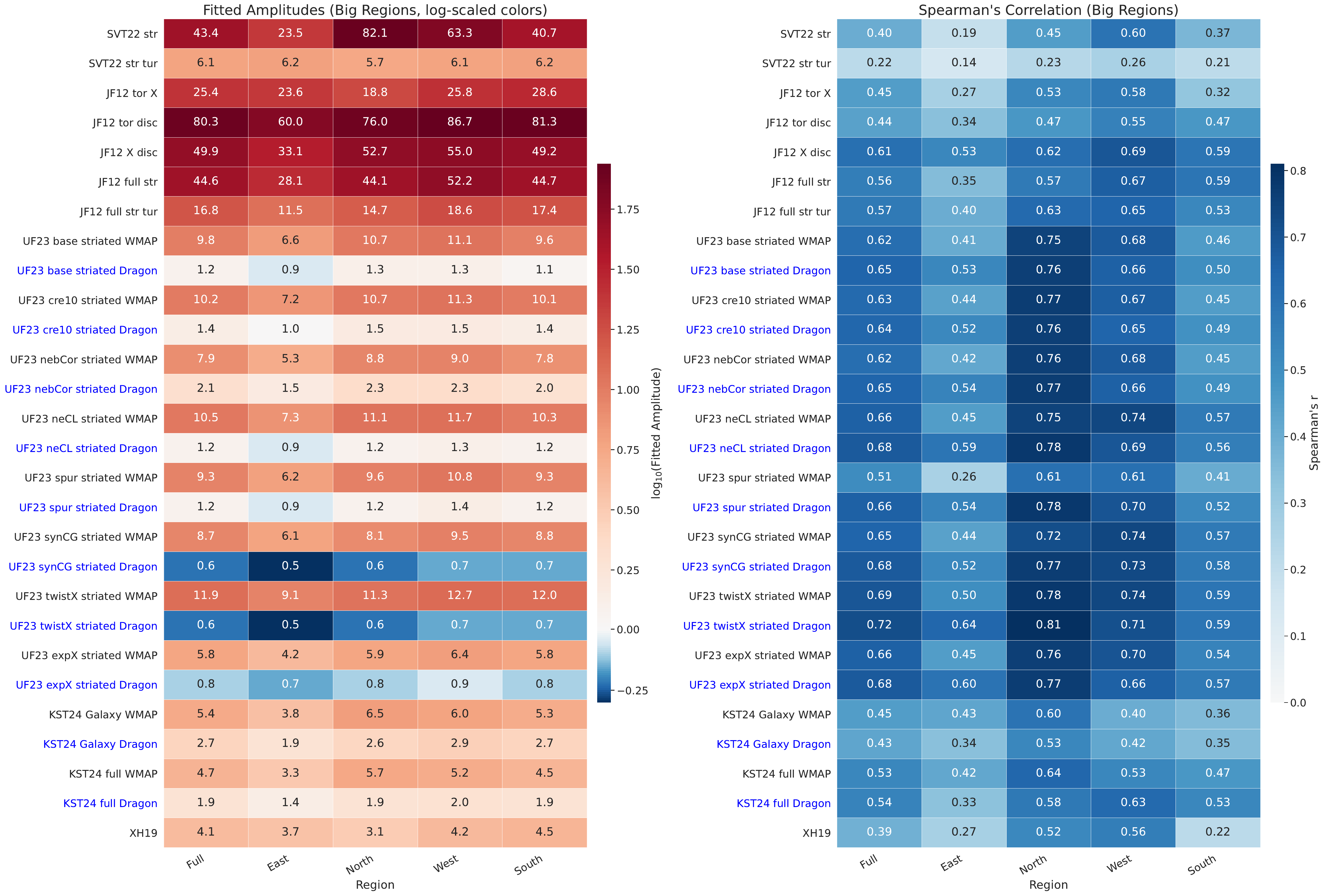}

\caption{{Heatmaps showing the fitted amplitude (left) and Spearman's rank correlation coefficient (right) for each GMF model across selected hemispheric regions, evaluated at HEALPix resolution \(N_{\mathrm{side}} = 16\). The fitted amplitude represents the scaling factor obtained from the linear template fit required to match the modelled polarized intensity to the observed C-BASS/S-PASS data at 4.76\,GHz. The Spearman's correlation coefficient quantifies the monotonic morphological agreement between model and data, independent of absolute amplitude. The colour scales and conventions are the same as in Figure~\ref{tab:amp_Spearman_final_8Quad}. Both metrics are computed regionally using the masks defined in Figure~\ref{fig:region_masks}. All skymaps are calculated using the \textit{WMAP} electron distribution model except \textit{KST24 full/Galaxy Dragon}.}}

    \label{tab:amp_Spearman_final_Big}
\end{figure*}

\begin{figure*}
    \centering
    \includegraphics[width=0.90\linewidth]{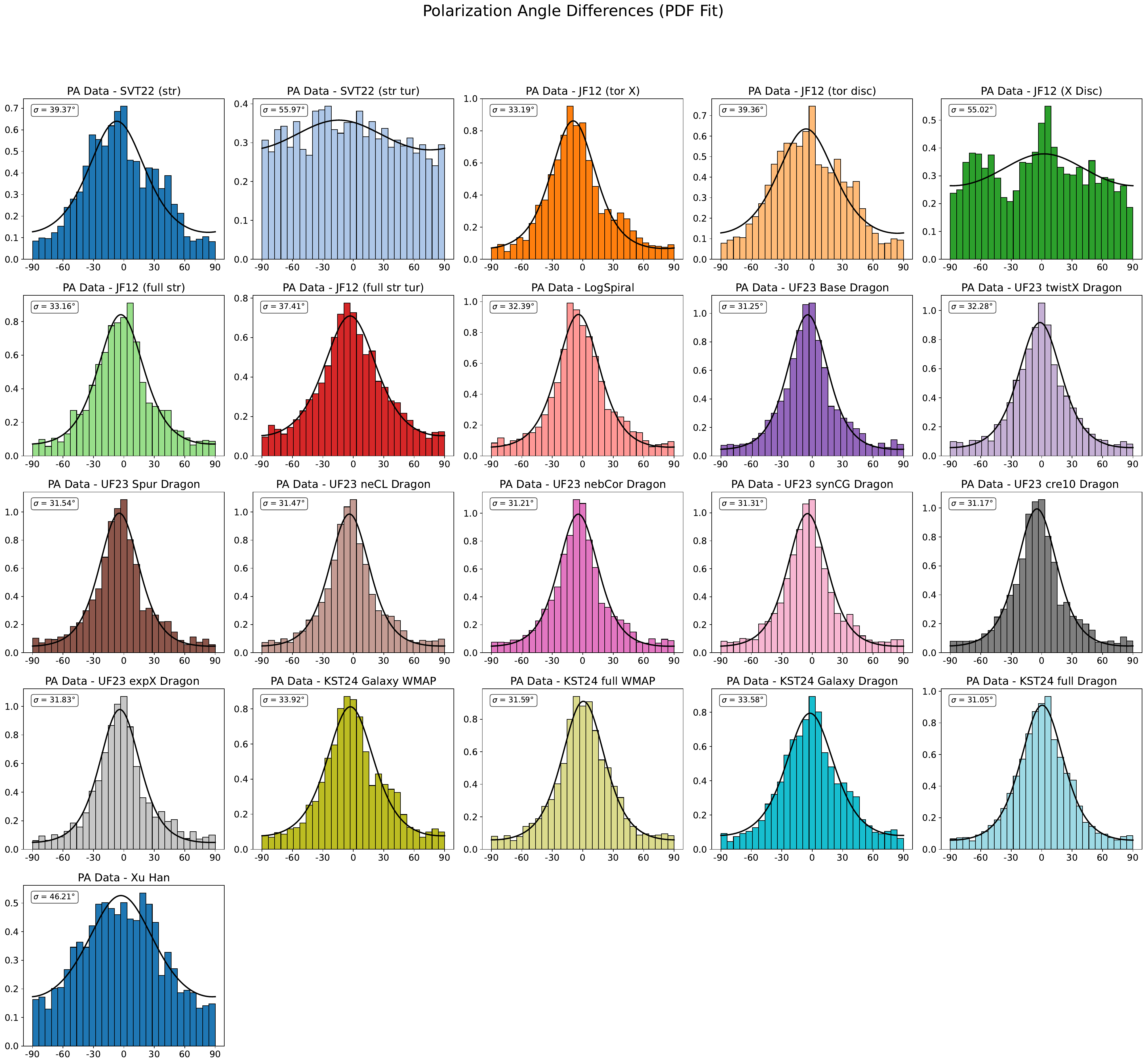}
    \caption{Histograms of polarization angle (PA) differences between the C-BASS/S-PASS data and each GMF model, computed at HEALPix resolution \(N_{\rm side} = 16\). The distributions are fitted using the position angle difference model described by \protect\cite{Khouei_1993}, which accounts for the circular nature of angle measurements. The standard deviation of each distribution is reported in the legend, serving as a quantitative measure of angular agreement. Lower standard deviation values indicate better alignment between the modelled and observed polarization angles.}
    \label{fig:pa_diff_histograms}
\end{figure*}
\subsection{Methods and overview}
We now make a quantitative comparison between the predictions of the synchrotron continuum emission from the aforementioned Galactic magnetic field models and the observations. We apply a template-fitting (or correlation) analysis, a well-established method for studying Galactic foregrounds \citep{Kogut_1996,Planck_XXI,Peel_2012,Planck_XXII,Harper_2022}. This involves modelling the sky signal $\textbf{d}$ as a linear combination of $N_{\mathrm Z}$ templates and adding a noise vector $n$:
\begin{align}
\textbf{d} = \textbf{Z}\textbf{a} + \textbf{n},
\end{align}
where each column of $\textbf{Z}$ ($N_{\mathrm {pix}} \times N_{\mathrm Z}$) represents the pixel values of a template. In our case, we use the modelled polarized synchrotron emission as the sole template. The coefficient vector $\textbf{a}$ is solved for via least squares as
\begin{align}
\hat{\textbf{a}} = \left(\textbf{Z}^{\mathit{T}} \textbf{N}^{-1} \textbf{Z}\right)^{-1} \textbf{Z}^{\mathit{T}} \textbf{N}^{-1} \textbf{d},
\end{align}
and \(\hat{\textbf{a}}\) represents the amplitude needed for the model to best match the observed 4.76\,GHz polarized synchrotron data.

We choose to use PI and polarization angle (PA) maps rather than Stokes \textit{Q} and \textit{U}, which are more coordinate dependent, because PI and PA correspond more directly to the strength and orientation of the magnetic field, respectively, making physical interpretation more intuitive.

\subsubsection{polarized intensity maps}
\label{sec:sec5_PI}
Figure~\ref{fig:PI_model_maps} presents PI maps of synthetic synchrotron emission from the 21 GMF models and model combinations. Each map has been scaled by the best-fit amplitude obtained from the full-sky template fit described above. The maps are shown in Galactic coordinates using a Mollweide projection, with masked regions such as the Galactic plane and areas contaminated by foregrounds indicated in grey. The colour scale is continuous and logarithmic, allowing detailed visualisation of the spatial variation in polarized intensity across the sky for each model. For clarity, the electron distribution model used to calculate each skymap in the case of the \textit{UF23} and \textit{KST24} models has been specified in the title of each PI map for the respective model. The PI skymaps for \textit{SVT22}, \textit{JF12} variants, \textit{LogSpiral}, and \textit{XH19} are calculated only with the \textit{WMAP} electron distribution model, as noted in  Section~\ref{sec:EdNdE_disc}, and thus there are no post-fix electron-distribution model names attached to them.

In terms of PI morphology, models like \textit{SVT22 str} and \textit{JF12 full str} fail to capture the observed structure of the synchrotron emission, even at high Galactic latitudes and toward the outer Galaxy (i.e. $|\ell| \geq 90^\circ$). \textit{SVT22 str tur} especially struggles due to the high levels of turbulence present in the model. Among all models, \textit{KST24 full Dragon} and the \textit{UF23} variants provide the best overall visual correspondence with the observed polarized synchrotron emission at 4.76\,GHz. This agreement is primarily attributed to the inclusion of local features in the modelling, notably the Local Bubble and the Fan region. The \textit{KST24 full Dragon} map is a combination of the \textit{KST24 Galaxy Dragon}, representing the large-scale GMF, and \textit{KST24 Bubble Dragon}, which models the synchrotron contribution from the Local Bubble. The inclusion of these components yields a morphology that more accurately reflects the observed emission across both high-latitude and outer Galactic regions.

\subsubsection{Residual polarized intensity maps} \label{sec:sec5_PI_residual}
Figure~\ref{fig:diff_PI_maps} presents residual PI  maps following amplitude rescaling using the best-fit coefficients obtained from the full-sky template fitting procedure. These residuals are computed as the difference between the observed and modelled synchrotron PI, with the colour scale indicating over-prediction (blue) and under-prediction (red) relative to the data.  Most residuals highlight the prominent NPS and the Fan region in red, as these are not incorporated in any of the models, except for the \textit{KST24} models which include the Fan region but not the NPS. Most of the \textit{UF23} variants calculated with \textit{Dragon}, and \textit{KST24 full Dragon} exhibit the low residuals across most of the unmasked sky, indicating a high degree of agreement with the observed PI structure. In particular, these models perform well in high-latitude regions and across broad angular scales, suggesting that their underlying magnetic field configurations and synchrotron emission modelling are more consistent with the data both morphologically and in amplitude.

\subsubsection{polarization angle maps} \label{sec:sec5_PA}
Figure~\ref{fig:pol_angles} presents the PA maps derived from the 21 GMF model combinations. In contrast to the PI maps, the PA distributions exhibit broadly consistent large-scale morphology across most models when compared to observational synchrotron data, with a few notable exceptions. For instance, the \textit{SVT22 str tur} model appears dominated by small-scale fluctuations, resulting in a noisy and incoherent angle pattern. Likewise, the \textit{JF12 X disc} model, lacking a toroidal magnetic field produces a morphology that departs significantly from the radio data, highlighting the necessity of combining both toroidal and poloidal field components to adequately reproduce the observed angular structure.
The importance of incorporating a poloidal (X-shaped) field is highlighted by models such as \textit{SVT22 str} and \textit{JF12 tor disc}, which deviate markedly from the radio data, particularly near Galactic longitude $0^\circ$. Although the inclusion of turbulent fields in \textit{JF12 str tur} improves the complexity of the model, it introduces too much small-scale noise that clearly does not agree with observations on finer angular scales. The overall similarity among the remaining models is likely due to the presence of both toroidal and poloidal field components in their construction, as is the case for several \textit{JF12}, \textit{UF23}, and \textit{KST24} variants.
\subsubsection{Comparing different regions of the sky between model and data}
\label{sec:sec5_comparison}

To quantitatively evaluate the agreement between the GMF model predictions and the observed synchrotron data, we performed a linear template fit across multiple sky regions, denoted as \textit{Q-NE, Q-1-NE, Q-2-NE, Q-NW, Q-3-NW, Q-4-NW, Q-SE, Q-1-SE, Q-2-SE, Q-SW, Q-3-SW, Q-4-SW, Full, East, North, West} and \textit{South} (see Figure~\ref{fig:region_masks} for a visualisation of the different regions). For each region, we computed the fitted amplitude representing the scaling factor required to match the model to the data, as well as the Spearman’s rank correlation coefficient, which captures the morphological agreement irrespective of absolute amplitude. {Figures~\ref{tab:amp_Spearman_final_8Quad} and \ref{tab:amp_Spearman_final_Big} present a summary of this analysis for the different regions and 
 serve complementary purposes. 
Figure~\ref{tab:amp_Spearman_final_8Quad} presents the fitted amplitudes and  correlation coefficients in smaller, morphology-driven sub-regions, enabling  a spatially resolved assessment of model performance and the impact of 
prominent local structures. In contrast, 
Figure~\ref{tab:amp_Spearman_final_Big} evaluates the same metrics over larger hemispheric regions, providing a broader-scale view of how each GMF model performs across the sky.}

In these figures, the left panel shows the fitted amplitudes, while the right panel shows the corresponding Spearman’s correlation coefficients for each model across all regions. Models are listed on the vertical axis, and sky regions along the horizontal axis. Notably, we include only “striated” versions of the \textit{UF23} models, where the striation factor $(1 + \beta)$ accounts for anisotropic turbulence aligned with the large-scale field. The inclusion of this factor leads to a marked improvement in the fitted amplitude values, often bringing them closer to unity, particularly in models such as \textit{UF23 neCL striated Dragon}, \textit{UF23 spur striated Dragon}, and \textit{UF23 twistX striated Dragon}.

From a quantitative standpoint, models such as \textit{SVT22 str}, \textit{SVT22 str tur}, and several \textit{JF12} variants (\textit{JF12 tor disc}, \textit{JF12 X disc}, etc.) require substantially higher fitted amplitudes across nearly all regions. This suggests that these models systematically underestimate the intrinsic synchrotron emission, likely due to insufficient magnetic field strength, the omission of key components, or an inaccurate electron distribution model (\textit{WMAP} in this case). The Spearman’s coefficients further support these findings: high values (typically $r > 0.6$) are observed for \textit{UF23} and \textit{KST24 full} across most regions, indicating good morphological agreement with the data irrespective of the electron distribution model adopted. In contrast, models with lower structural fidelity, such as \textit{SVT22 str tur} or \textit{JF12 X disc}, exhibit weak or even negative correlations in some regions, especially those sensitive to local structures (e.g. \textit{Q-2-SE, North}).

\subsubsection{Comparison of polarization angle difference distributions between models and observations}

In Figure~\ref{fig:pa_diff_histograms}, we present histograms showing the difference in polarization angles between the C-BASS data and the predictions from various GMF models. These histograms quantify the angular offset between the observed synchrotron polarization angles and those predicted by each model. Each histogram is fitted using the angle-difference distribution described in Equation (3) of \cite{Khouei_1993}, with the corresponding standard deviation (\( \sigma \)) shown in the legend of each panel. The value of \( \sigma \) serves as a proxy for the model's angular agreement with the data; smaller \( \sigma \) indicates a tighter match. 
    
Among the models, \textit{UF23 Base WMAP}\footnote{Note: the striation factor for \textit{UF23} models is a constant value and only affects the polarized intensity amplitude and thus has no effect on polarization angles.}, \textit{UF23 cre10 WMAP}, \textit{KST24 full WMAP/Dragon}, and \textit{UF23 nebCor WMAP} exhibit the smallest standard deviations, reflecting strong alignment with the C-BASS polarization angles. In contrast, models such as \textit{SVT22 str tur} show significantly broader distributions. The high standard deviation in this model arises from the strong turbulent component, which introduces substantial small-scale fluctuations and noise, resulting in an almost flat distribution of angle differences. These results reinforce the utility of angle-difference histograms as a diagnostic for assessing GMF model performance, especially in capturing the coherent large-scale structure of the GMF.

Below we discuss in detail each model.
\subsection{Performance of individual models calculated with the analytical electron distribution}

\subsubsection{SVT22 model}
The \textit{SVT22 str} and \textit{SVT22 str tur} models consistently underperform in reproducing the observed features in both polarized intensity (PI) and polarization angle (PA). Of all models examined, \textit{SVT22 str} provides the poorest fit, requiring high amplitude scaling factors in most sub-regions and exceeding 100 in regions such as \textit{Q-2-NE} and \textit{Q-2-SE}. While performance improves slightly for larger hemispherical regions (e.g. \textit{North}, \textit{South}, \textit{East}, \textit{West}), the minimum required amplitude remains high, with the \textit{East} hemisphere still demanding a factor of 24.

This performance can be attributed to the model's lack of a poloidal magnetic field component, which limits its ability to reproduce synchrotron features at higher Galactic latitudes and in the outer Galaxy. Additionally, it does not account for local structures such as the Fan region or the Local Bubble.

Although the amplitude values fitted to the \textit{SVT22 str tur} model may appear reasonable in some areas, they are not physically meaningful. The low Spearman's correlation coefficients indicate a lack of morphological agreement with observations, suggesting that the model is primarily fitting to noise fluctuations rather than coherent synchrotron structures.

For polarization angle, the \textit{SVT22 str} model may appear superficially aligned with observations at large angular scales. However, a closer inspection reveals a lack of coherence, especially in the outer Galaxy, again due to the missing poloidal component. The \textit{SVT22 str tur} model performs more poorly: its PA maps are dominated by turbulent noise and show no coherent structure.

The PA difference maps and histograms further confirm these shortcomings. The \textit{SVT22 str tur} model yields the highest dispersion ($\sigma \approx 56^\circ$) in angle differences, with the PA difference histogram being essentially constant, indicating no correlation in angle, while \textit{SVT22 str} also performs poorly with $\sigma \approx 40^\circ$. These large dispersions, together with the morphology and amplitude mismatches, clearly demonstrate the inadequacy of both \textit{SVT22} variants in capturing the observed synchrotron polarization.

\subsubsection{JF12 model}
We evaluate a set of structured and turbulent magnetic field variants of the \textit{JF12} model to assess the impact of different magnetic field geometries on the polarized synchrotron emission. While several of the configurations analysed do not correspond exactly to the full original model, their inclusion allows for a more systematic investigation of the role of individual field components in shaping the synchrotron morphology.

The \textit{JF12 full str} model requires relatively large fitted amplitudes but maintains consistent correlation with the data across most regions. A notable exception is region \textit{Q-2-NE}, where the model exhibits negative correlation. This behaviour is attributable to a sharp drop in synchrotron brightness beyond Galactic longitude $90^{\circ}$, caused by the abrupt termination of the toroidal field in this configuration. On larger scales, the model performs least effectively in the \textit{East} hemisphere, where the poloidal component dominates over the toroidal field both above and below the masked Galactic plane, leading to suppressed brightness in regions beyond $90^{\circ}$.

The best-performing configuration among the \textit{JF12} variants is \textit{JF12 full str tur}, which includes both coherent and turbulent magnetic field components. This model requires lower fitted amplitudes and gives higher Spearman's correlation coefficients across most regions, with particularly strong performance in region \textit{Q-4-NW}. Despite this, none of the \textit{JF12} configurations are able to reproduce key morphological features observed in the PI maps, such as the NPS and Fan region, implying that an additional model for such regions is necessary.

The \textit{JF12 tor X} model yields fitted amplitudes that are higher than those of the complete structured model (\textit{JF12 full str}); however, the associated Spearman’s correlation coefficients for the former are notably lower. This suggests that, although the global intensity scaling may be consistent, the spatial agreement with the data is diminished in the absence of the disc field. Despite its limited contribution at high latitudes, the disc component appears to play a key role in enhancing the morphological coherence in regions near the Galactic plane, thereby contributing significantly to the overall correlation.

Similarly, the \textit{JF12 X disc} configuration exhibits systematically high fitted amplitudes across all sky regions, yet maintains average Spearman’s coefficients around 0.6, higher than those found for \textit{JF12 tor X}. This discrepancy can be attributed to the nature of Spearman’s rank correlation coefficient, which measures the degree of monotonic association between the model and data. Consequently, even in cases where the morphological agreement is limited, a high coefficient can result if the relative ordering of pixel intensities is preserved.

The PA maps from most \textit{JF12} configurations exhibit broad agreement with the observed data on large angular scales. A clear exception is the \textit{JF12 X disc} model, whose lack of a toroidal component results in a PA morphology that deviates substantially from observations. The \textit{JF12 full str} and \textit{full str tur} configurations provide the best agreement in PA structure, although the latter displays enhanced small-scale fluctuations due to the added turbulent component, which are not supported by the data.

This behaviour is further confirmed by the PA difference maps and histograms. The inclusion of turbulence in \textit{JF12 full str tur} reduces the angular dispersion in several regions, yet significant deviations remain. The corresponding angular standard deviations are typically in the range $\sigma \approx 30^\circ$–$35^\circ$, indicating improved but still incomplete agreement with the observed polarization angle patterns.

\subsubsection{UF23 models}
Each configuration within the \textit{UF23} model family includes an optional striation factor provided in Table~3 in \citep{Unger_2024}, designed to account for the contribution of anisotropic turbulent (or striated) magnetic fields. The inclusion of this factor systematically improves model performance across all sky regions, resulting in lower fitted amplitudes. The Spearman’s correlation coefficients remain unaffected by the striation factor, as the striation factor is merely a constant value. We show results only for the striated versions of the \textit{UF23} models in Figures~\ref{tab:amp_Spearman_final_8Quad} and \ref{tab:amp_Spearman_final_Big}.
 
Among the various configurations, the \textit{UF23 twistX striated WMAP} model requires the highest fitted amplitude but simultaneously achieves one of the highest correlation coefficients. This may be attributed to the complex twisted halo field geometry arising from large-scale plasma flows in the Galactic halo. 

In addition, models such as \textit{UF23 neCL striated Dragon}, \textit{UF23 spur striated Dragon}, and \textit{UF23 synCG striated Dragon} also perform well, with fitted amplitudes fluctuating around a factor of 1–2 and Spearman’s coefficients consistently exceeding 0.6 in most regions, with the exception of \textit{Q-SE}, \textit{Q-1-SE}, and \textit{Q-2-SE}. The \textit{WMAP} counterparts for these models require higher fitted amplitudes and have lower Spearman’s coefficients. This discrepancy between the electron distribution models is discussed in detail in Section~\ref{sec:diff_EdNdE_disc}. 

The highest correlation values across nearly all model variants are observed in region \textit{Q-NW}, which is likely dominated by diffuse synchrotron emission from the large-scale GMF, and thus less sensitive to localized structures.  {Although none of the \textit{UF23} models explicitly incorporate a dedicated component for the Fan region,
morphological features qualitatively resembling the Fan can nevertheless be identified in most model variants, albeit with varying levels of brightness and prominence.}

The PA maps produced by the \textit{UF23} models are broadly consistent across different configurations and align well with observational data on large angular scales. The corresponding angular dispersions, quantified by the standard deviation $\sigma$ of the PA differences, are generally lower for these models, typically in the range $\sigma \approx 30^\circ$, and indicate improved agreement with the data, particularly in regions dominated by large-scale magnetic structure. As noted previously, the striation factor does not affect the PA maps or the histogram of the PA differences.

\subsubsection{KST24 models}
The \textit{KST24 full WMAP} model incorporates both a large-scale GMF component and a localized contribution from the Local Bubble and the Fan region, used in calculating the \textit{WMAP} electron distribution model. It is the only model in this analysis that explicitly reproduces the Fan region in the polarized intensity (PI) maps. However, it does not capture the morphology of the NPS, which remains a persistent challenge for Galactic synchrotron modelling. The fitted amplitude for this model lies mostly between 1–2, with the \textit{West} hemisphere being a perfect fit to the data, which is most likely because this region does not have any known local foregrounds in the data. With regard to the Spearman’s coefficient, the model performs best in \textit{Q-4-NW} and \textit{Q-4-SW}, both being part of the \textit{West} hemisphere.

To better understand the contribution of the individual components, we also examine the performance of the \textit{KST24 galaxy WMAP} model separately from the full model. In contrast, the \textit{KST24 galaxy WMAP} model exhibits more variation, with fitted amplitudes between 3–7 throughout the sky. For the Spearman’s coefficient, most regions have a moderate value of 0.5; however, regions \textit{Q-SE} and \textit{Q-SW} perform poorly, most likely due to the lack of the Bubble model. 

With respect to polarization angle (PA), both the \textit{KST24 full WMAP} and \textit{KST24 galaxy WMAP} models show broad agreement with the data at large angular scales. The inclusion of the Local Bubble has a noticeable effect on the angular distribution, improving the local coherence and enhancing the alignment of model and data in several regions. This is evident in the PA difference maps, where the full model exhibits smaller angular deviations across much of the sky. The \textit{KST24} model includes the Fan region, for PA there is a clear mismatch in angles below $10^\circ$ latitude in the southern hemisphere between the models and the data.

The \textit{KST24 full WMAP} model yields a lower angular dispersion, with a typical value of $\sigma \approx 32^\circ$, compared to the \textit{KST24 galaxy WMAP} variant, which shows slightly larger values of $\sigma \approx 34^\circ$. These results highlight the significant impact of localized magnetic structures, such as the Local Bubble, in shaping both the intensity and angle morphology of the polarized synchrotron sky. 

We discuss the effect of the \textit{Dragon} electron distribution model for the \textit{KST24} models in Section~\ref{sec:diff_EdNdE_disc} and compare it with the \textit{WMAP} maps. 

\subsubsection{XH19 model}
The \textit{XH19} model is characterised by an asymmetric toroidal halo field, which results in a sky map that appears notably asymmetric between the northern and southern hemispheres. This large-scale structural asymmetry contributes to a lack of overall agreement with the observed polarized synchrotron data.

Across all sky sub-regions, the model exhibits consistently weak correlations, particularly in regions \textit{Q-2-NE}, \textit{Q-2-SE}, and \textit{Q-3-SW}, where the model and data are anti-correlated. This is reflected in both the fitted amplitude maps, which show high rescaling factors in these regions, and the difference maps, which highlight strong morphological mismatches in the polarized intensity (PI) structure.

The model's polarization angle (PA) morphology also deviates significantly from observations. Owing to its predominantly toroidal configuration and the absence of a poloidal or localized component, the resulting PA maps lack the coherent angular structures present in the data, particularly at high latitudes. This discrepancy is evident in the PA difference maps, which show widespread angular misalignments.

The severity of this misalignment is quantified by the residual angle distribution, where the \textit{XH19} model yields one of the largest angular dispersions among all models considered, with a standard deviation of approximately $\sigma \approx 47^\circ$. This large value underscores the model’s inability to reproduce the observed polarization angle patterns, further highlighting the limitations of relying on a purely toroidal magnetic field structure.

\begin{figure*}
    \centering

    \includegraphics[width = 0.49\linewidth]{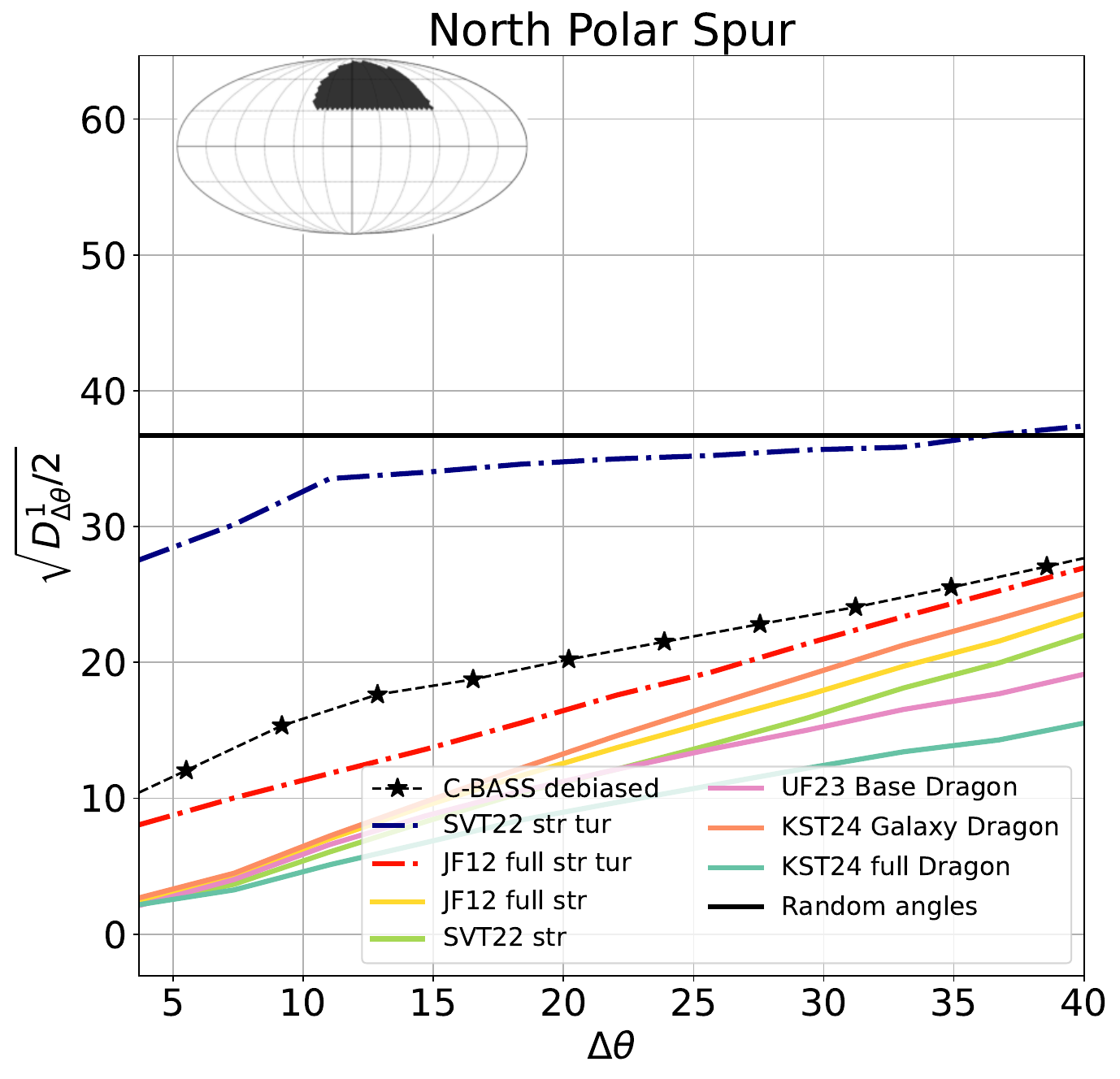}
    \includegraphics[width = 0.49\linewidth]{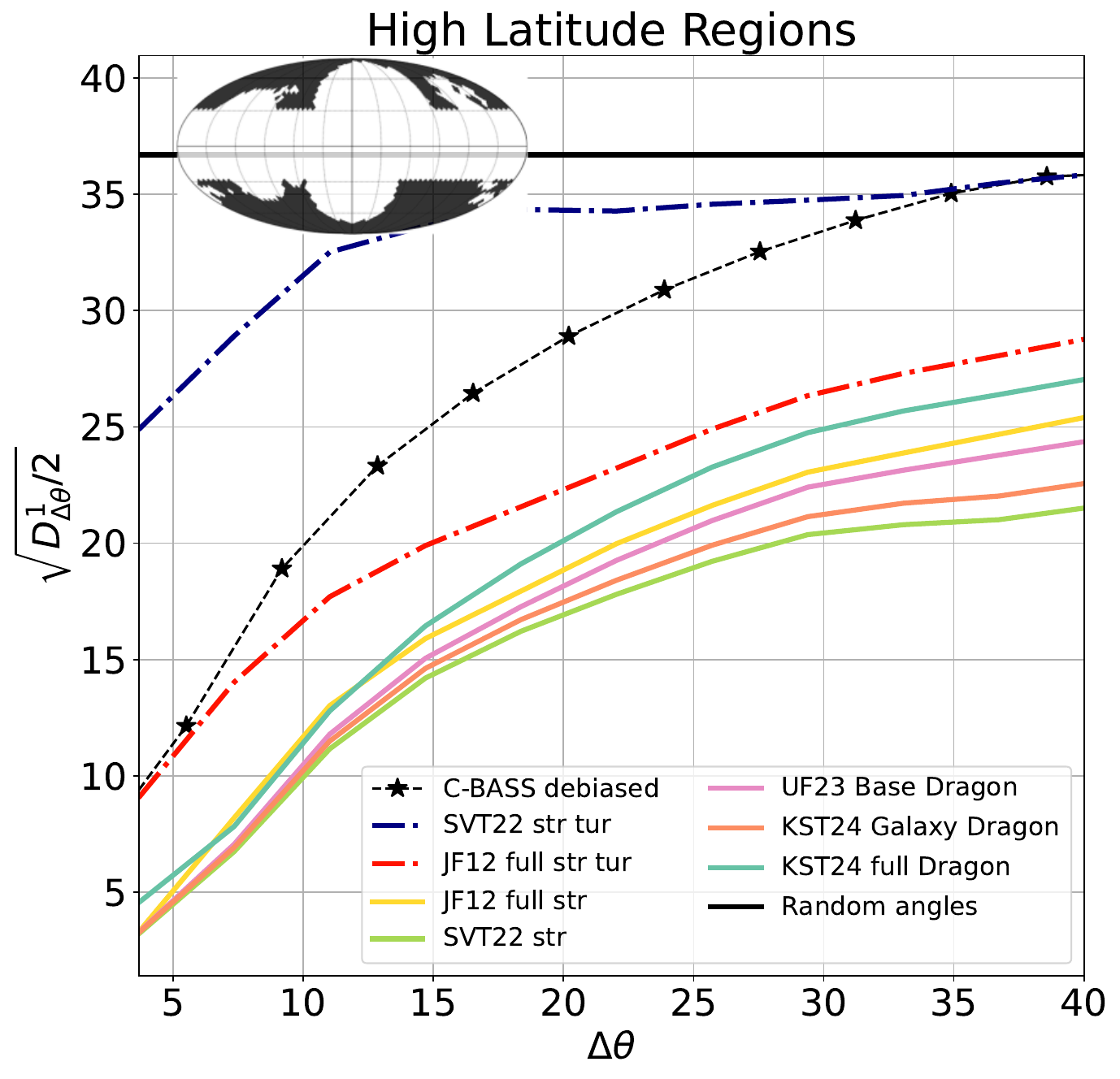}

    \caption{Polarization angle structure functions at {$N_\mathrm{SIDE}=16$} ( $\approx  {3.7^\circ}$) for the North Polar Spur and high Galactic latitude regions of the reconstructed 4.76\,GHz sky. The structure functions were calculated for the reconstructed 4.76\,GHz sky and compared with polarization angles obtained from the \textit{JF12 full str tur, JF12 full str}~\citep{JF12,Jansson_2012}, \textit{SVT22 str tur, SVT22 str}~\citep{Shaw22}, 
    \textit{UF23 Base Dragon}~\citep{Unger_2024}, and \textit{KST24 full Dragon, KST24 Galaxy Dragon}~\citep{Korochkin_2025} models along with a random angle map. {The solid black line is obtained from the random angle map.} }

    \label{fig:stfunc}
\end{figure*}

\section{Structure Functions}
\label{sec:strucfunc}

Structure functions are powerful tools for quantifying variability in astronomical data. First introduced in radio astronomy to study interstellar scintillation \citep{Rickett1977}, they have since been applied to analyse variability in extragalactic radio sources, proving effective even with irregular sampling \citep{Simonetti1984,Simmons1985,Minter1996,Haverkorn2004}. In polarization studies, structure functions quantify the coherence of polarization angles across angular scales. They have been used to study time variability in sources such as pulsars \citep{Lam2020} and the spatial coherence of rotation measure in radio galaxies \citep{Leahy1986,Leahy1987}. Unlike Fourier-based methods, structure functions are well-suited for datasets with incomplete sky coverage, such as the \mbox{C-BASS} map or heavily masked regions. In this work, we use the second-order structure function to measure polarization angle coherence, avoiding power spectrum analysis due to the complexity of masking and the associated window functions.

The second-order structure function is defined as:
\begin{equation}
D^2(\Delta{\theta}) = \left \langle \left [ \chi^\prime(\theta+\Delta\theta)-\chi^\prime(\theta) \right ]^2 \right \rangle,
\end{equation}
where $\chi^\prime(\theta)$ is the polarization angle at position $(\ell,b)$ and $\chi^\prime(\theta+\Delta\theta)$ is its value at an angular separation $\Delta\theta$. The quantity $\sqrt{D^2(\Delta\theta)/2}$ corresponds to the RMS of angle differences at scale $\Delta\theta$. The structure function also relates to the variance $\sigma^2$ and the autocorrelation function $\rho(\theta)$:
\begin{equation}
D^2(\Delta\theta) = 2\sigma^2\cdot\left [ 1-\rho(\Delta \theta) \right ],
\end{equation}
valid for scalar quantities when both $\rho(\theta)$ and $D^2(\Delta{\theta})$ are evaluated over a region much larger than the outer scale of the correlation. (Unlike $\rho(\theta)$, $D^2(\Delta{\theta})$ is not a biased estimator when this condition fails to hold.) The angle differences are folded into $\pm 90^{\circ}$ to preserve the $180^{\circ}$ cyclic property of polarization angles.

To make $D^2(\Delta{\theta})$ independent of the spherical coordinate system, we compare position angles in different directions using parallel transport. We compute the difference in angles using the $\chi^\prime$ polarization angles instead of the raw $\chi$ polarization angles:
\begin{equation}
\chi^\prime(\theta+\Delta\theta)-\chi^\prime(\theta) = \left [ \chi(\theta+\Delta\theta)-\alpha_\mathrm{final} \right ] - \left [ \chi(\theta) - \alpha_\mathrm{initial} \right ],
\end{equation}
where $\alpha_\mathrm{initial}$ and $\alpha_\mathrm{final}$ are the initial and final headings along the great-circle segment between the two sky coordinates.

Structure functions are computed for high-latitude regions in both the northern and southern hemispheres, using maps degraded to $N_\mathrm{side} = 16$ to ensure quasi-independent sampling. The structure function analysis is constrained at large angular scales by the finite sky coverage and map edges; therefore, we restrict our analysis to smaller angular scales. As the structure function is invariant under global rotations, absolute polarization angle calibration does not affect the results. {Following \citet{Haverkorn2004}, the error bias has been subtracted, and pixels with polarization angle errors greater than $10^\circ$ are excluded, removing points where the PA error distribution is substantially non-Gaussian. The overall correction is very small due to the 
high signal-to-noise ratio in most pixels at a resolution of $N_\mathrm{side} = 16$ ($\sim 3.7^{\circ}$ pixel size), thereby ensuring a high $S/N$ sample. All remaining angles are then equally weighted, so that the results are not biased towards regions with higher $S/N$. The angles were not Faraday corrected to zero wavelength as at these high latitudes the corrections are small and their uncertainties are large.}

We estimate the structure functions\footnote{We tested a random noise map with uniform random angles, both with and without masking, and obtained a straight line at a level coinciding with the expected value of the RMS up to the largest angular scales allowed by the mask. Additionally, our code was tested on different projections of the sky, such as Galactic and Ecliptic, to confirm that the results remain invariant under rotations.} for the reconstructed 4.76~GHz sky polarization angle data and compare them with those from four GMF models: \textit{JF12} \citep{JF12,Jansson_2012}, \textit{SVT22} \citep{Shaw22}, \textit{UF23 Base} \citep{Unger_2024}, and \textit{KST24} \citep{Korochkin_2025}, both with and without the inclusion of turbulent components. The analysis focuses on selected regions, including the North Polar Spur and high-latitude sky in both hemispheres, as defined by the applied mask shown in the top-left panel of Figure~\ref{fig:stfunc}. As a reference, a map of random polarization angles for $\delta > -15^\circ$ is included, yielding a flat structure function with $\sqrt{D^2(\Delta\theta)/2} = 90^\circ/\sqrt{6} \approx 36.7^\circ$, consistent with the expectation for uncorrelated angles.

Only scales of up to $40^\circ$ are shown in Figure~\ref{fig:stfunc}, since the structure functions become unrepresentative for separations near the size of the mask. For example, in the case of the NPS, the smallest region shown, larger angular scales ($60^\circ$–$90^\circ$) are derived only from differences between pixels near opposite edges of the mask. While smaller angular scales are more easily interpreted and have even spatial coverage, the largest scales often represent particular parts of the sky near the edge of the mask.

The structure functions all increase with angular scale, implying that larger scales are less coherent than smaller scales, as expected. The structure function from the 4.76~GHz data (shown by the black line on the left-hand side in Figure~\ref{fig:stfunc}) is coherent and lies below the random-angle level for the NPS case, implying that correlations in the polarization angles exist in this region. Some models, such as \textit{JF12 str tur}, still yield polarization angles that align relatively well with the observed angles in the NPS region. This may be coincidental and could arise from large-scale toroidal fields in the models producing a polarization angle distribution similar to that observed. Notably, the NPS region is known to be highly polarized—up to approximately {40\%–50\% after background subtraction is considered \citep{Vidal2015,Dickinson2018}; Cepeda-Arroita et al. (in preparation)—implying that a single coherent structure dominates. Whether this structure is local or Galacto-centric is still debated \citep{Panopoulou2021,eROSITA,Zhang_2024}. Thus, while the apparent agreement between model predictions and observations in the NPS may be coincidental, further investigation is required to confirm the underlying cause of this correlation. The \textit{KST24 full Dragon} model performs the worst in the NPS region, likely due to the combined effect of its Galaxy and Bubble components, which do not align with the data. In contrast, \textit{SVT22 str tur} shows the fastest decorrelation in the NPS region: although the toroidal component of the field projects into the mask and provides some correlation, the strong turbulent component dominates and quickly erases coherent structure.}

In the high-latitude case, the introduction of the southern hemisphere through the S-PASS data results in decorrelation between angles at large angular scales.\footnote{See Cepeda-Arroita et al. (in preparation) for a detailed discussion of structure functions for C-BASS-only data.} The structure functions for the high-latitude case from \textit{JF12 full str tur}, \textit{KST24 (Galaxy/full) Dragon}, \textit{UF23 Base Dragon}, and \textit{SVT22 str} all remain well below the random-field limit, indicating some correlation in the polarization angles. In \textit{SVT22 str tur}, however, the decorrelation is too large at small scales for the high-latitude regions as well.

\textit{JF12 full str tur} performs best in both the high-latitude and NPS regions, with particularly strong agreement in the high-latitude region below $\Delta \theta = 10^\circ$. This may suggest that different levels or characteristics of the turbulent magnetic field are at play in different parts of the Galaxy. The \textit{SVT22} and \textit{JF12} models used here employ Kolmogorov $5/3$-type turbulence, which requires minimum ($L_{\rm{min}}$) and maximum ($L_{\rm{max}}$) wavelengths to generate the turbulent field, along with the field amplitude. In the case of \textit{SVT22 str tur}, the value of the field amplitude is too high to match the observations, whereas in the case of \textit{JF12 str tur}, the value of $L_{\rm{max}}$ is likely too low beyond $\Delta \theta = 10^\circ$, where it starts to deviate significantly from the data. The structure function results show that setting the correct $L_{\rm{max}}$ value in the case of small-scale Kolmogorov-type turbulence plays a crucial role in the overall morphology of polarization angles.

\begin{figure*}
    \centering
    \includegraphics[width=0.49\linewidth]{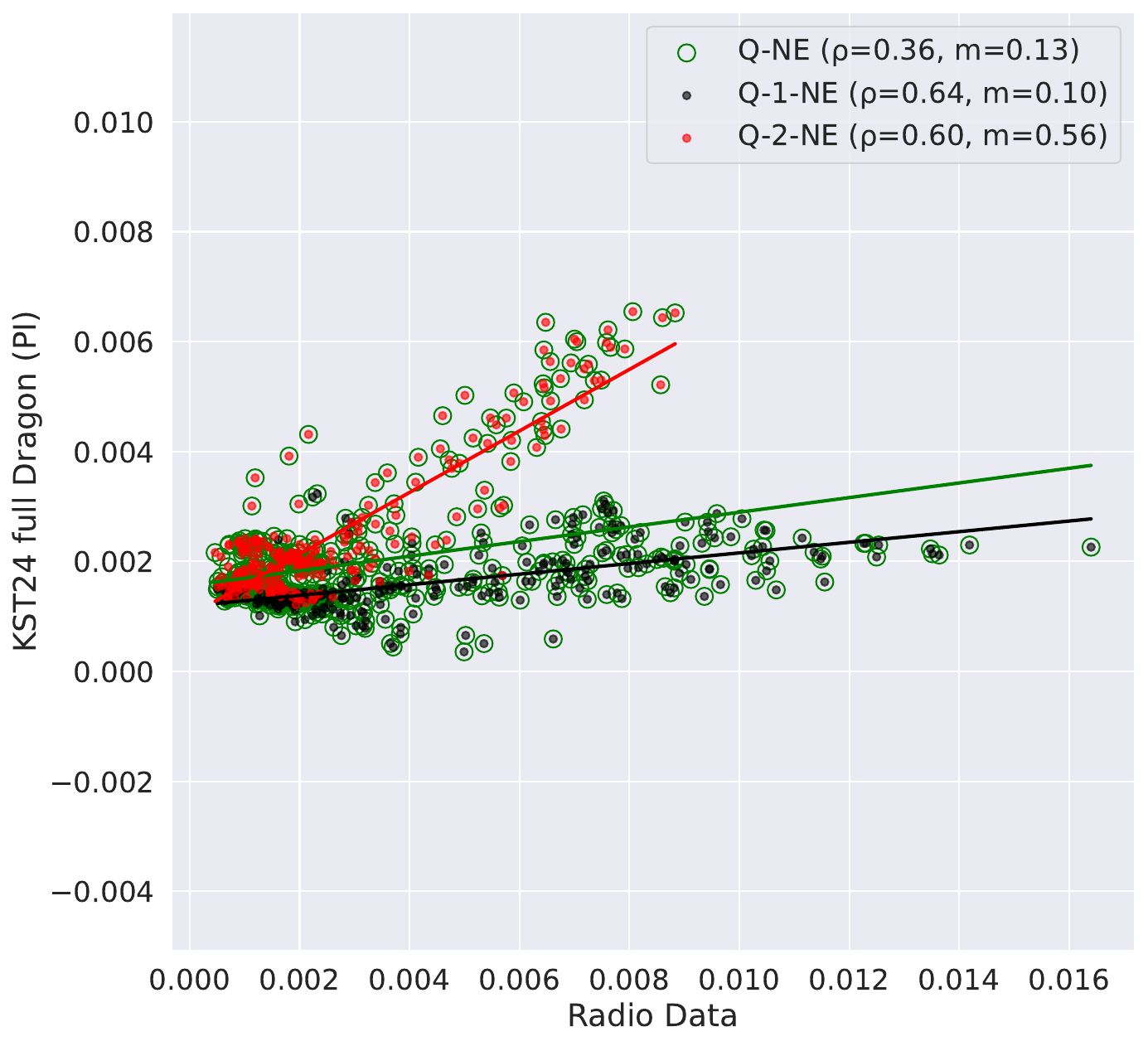}
    \includegraphics[width = 0.49\linewidth]{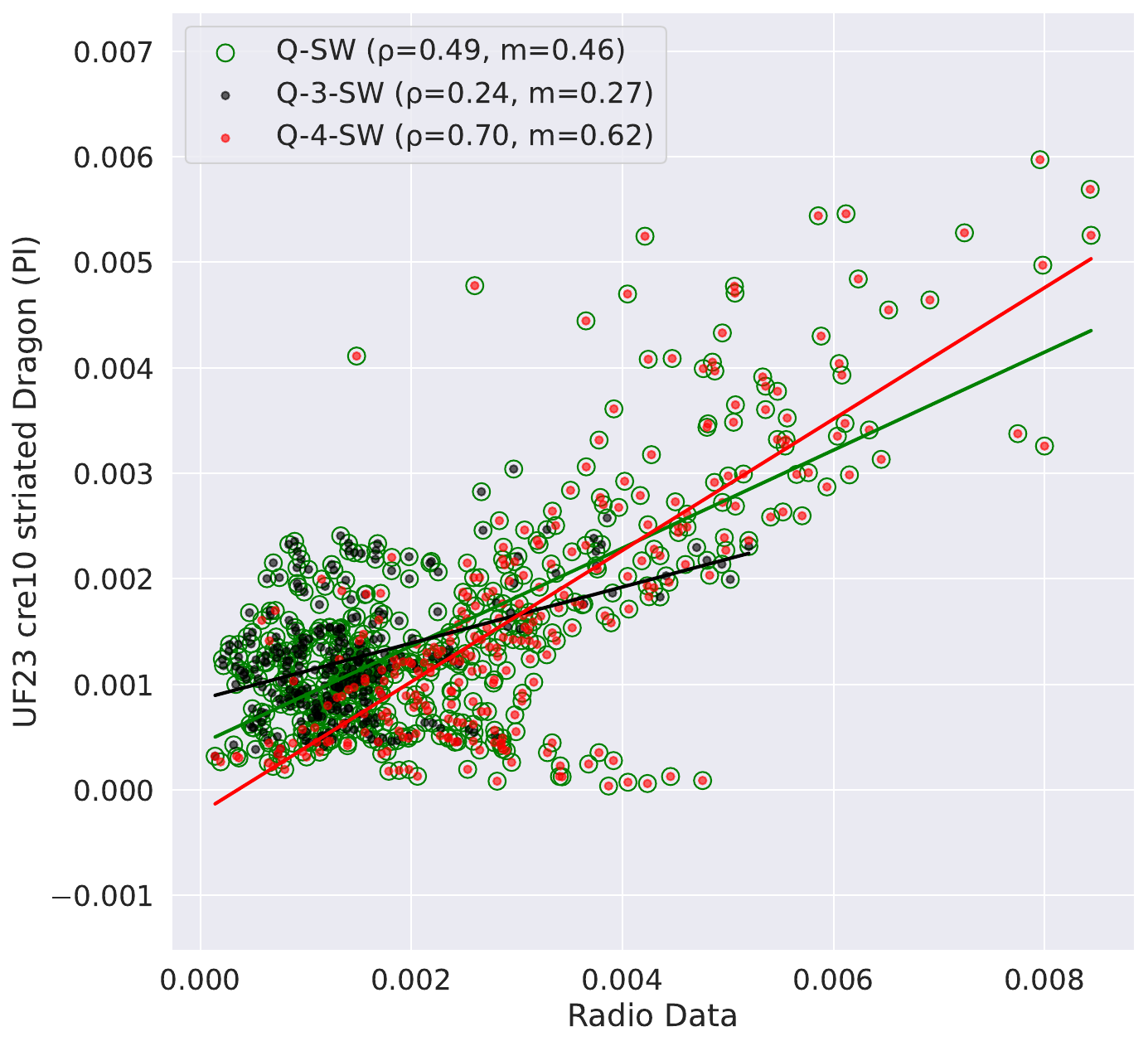}
 
    \caption{\textbf{Left :} T-T plot comparison between the data and the \textit{KST24 full \textit{Dragon}} model for regions\textit{Q-NE, Q-1-NE, Q-2-NE}.
    \textbf{Right:} T-T plot between the data and the \textit{UF23 cre10} model for regions\textit{Q-SW, Q-3-SW, Q-4-SW}.}
    \label{fig:tt-kst24-Q-NE_uf23cre10-Q-SW}
\end{figure*}

\section{Discussion}
\label{sec:discussion}

In this work, we present the first all-sky map of polarized synchrotron emission at 4.76~GHz, constructed by combining data from the C-BASS and S-PASS surveys. We use this map to perform a comprehensive large-scale comparison using template fitting with state-of-the-art GMF models and their component combinations, as summarized in Figures~\ref{tab:amp_Spearman_final_8Quad} and \ref{tab:amp_Spearman_final_Big}.

\subsection{Best-fit models and foreground components}

Our analysis reveals that most GMF models are unable to fully reproduce the observed brightness of the polarized synchrotron sky at 4.76~GHz without requiring an amplitude scaling. We divided the sky into multiple regions and identified which regions primarily drive this amplitude correction, from \textit{Q-NE} through \textit{Full} sky, and apply a template-fitting scheme to determine the fitted amplitude in each. We find that certain regions consistently require higher amplitudes across all models. Among these, some models perform better in the template-fitting analysis due to specific choices in their large-scale field components, such as \textit{UF23 nebCor Dragon} and \textit{UF23 neCL Dragon} \citep{Unger_2024}, or due to the incorporation of local Galactic structures, as in the case of the \textit{KST24 full \textit{Dragon}} model \citep{Korochkin_2025}. In terms of morphology, the \textit{KST24 full \textit{Dragon}} model is the only model that reproduces the brightness enhancement in the Fan region; however, it still under-predicts the emission in the NPS, likely due to the lack of any modelling of the magnetic fields in the NPS region. Some of the other models, such as \textit{SVT22 str} and \textit{XH19}, or individual components (e.g. \textit{JF12 tor disc, JF12 X disc}), fail to match the observed morphology, especially in regions dominated by local or non-axisymmetric features. The noise-dominated \textit{SVT22 str tur} requires significantly larger amplitudes, indicating low compatibility with the observed sky at high Galactic latitudes. A key conclusion of this study is that most GMF models, by neglecting local structures, are unable to reproduce the observed synchrotron emission. This suggests that the 4.76~GHz data are dominated by synchrotron radiation originating from nearby foreground features.

When it comes to polarization angles (PA), most models appear to reproduce the large-scale angular patterns seen in the data. This is particularly true for models that include multiple halo components, such as \textit{JF12 tor X}, \textit{JF12 full str}, and the \textit{UF23} models. The \textit{LogSpiral} model only matches the data in PA and we do not compare it in PI. The observed agreement in angles may not reflect a true spatial correlation. Instead, it is possible that the alignment arises by coincidence, without underlying spatial correlation, due to the geometry of the model.

{
Our finding that model–data discrepancies are region dependent may be
related to the influence of nearby magnetic structures, as suggested by
recent observational studies. \citet{West_2021} propose a unified
interpretation of the Fan region and the North Polar Spur (NPS) as
long, magnetized filamentary structures associated with the Local
arm and/or Local Bubble, emphasising that local emission can shape
large-scale polarized features. Likewise, \citet{Dickey2022} show that
large-angular-scale rotation-measure patterns are modulated by nearby
structures, including clear RM signatures associated with the
Orion–Eridanus superbubble, while \citet{Booth2025} demonstrate that
much of the observed Faraday sky within $\sim$1\,kpc of the Sun can
be described by a local magnetic field reversal geometry. Large-scale
structures associated with Galactic-centre activity, such as the
Fermi bubbles and the microwave haze
\citep{FERMI_2010,Fermi_Haze_2010,Planck_Haze_2013,Planck_X},
provide additional examples of non-axisymmetric features that may
influence the polarized sky at intermediate and high latitudes.

The \textit{KST24} model explicitly incorporates nearby components,
including the Local Bubble and the Fan region, and its comparatively
improved performance in certain regions is therefore potentially
suggestive that such local magnetic configurations may contribute
to the observed 4.76\,GHz polarized morphology.

In particular, regions \textit{Q-1-NE} and \textit{Q-2-NE} overlap
with the Fan region and Loop\,I/NPS, while region \textit{Q-SW}
overlaps with the Orion–Eridanus superbubble. Beyond these, Loop\,III
extends across parts of the southern sky and overlaps partially with
\textit{Q-3-SW} and \textit{Q-4-SW}, and Loop\,II lies within the
second Galactic quadrant, contributing to structure in
\textit{Q-2-NE}. Although Loop\,II is generally weaker in polarized
intensity than Loop\,I at 4.76\,GHz, it may still affect intermediate-
latitude morphology. 

While dominance in rotation measure does not necessarily imply
dominance in polarized intensity, the presence of multiple nearby
magnetized structures together with large-scale outflow features from
the Galactic Centre, could plausibly influence the regional correlation
behaviour and morphological discrepancies observed in our analysis. 
}

\subsection{Comparison of different cosmic ray electron distribution models: \textit{KST24} and \textit{UF23}}
\label{sec:diff_EdNdE_disc}

We compare the synchrotron emission obtained using the \textit{WMAP} analytical cosmic ray electron distribution and the \textit{Dragon} simulation for the \textit{KST24 full} and \textit{UF23} magnetic field models. In Figs~\ref{tab:amp_Spearman_final_8Quad} and \ref{tab:amp_Spearman_final_Big}, results using the \textit{Dragon} model are shown in blue, and those using \textit{WMAP} are shown in black. For the \textit{WMAP} model with parameters $R_{\rm el}=5\,\mathrm{kpc}$ and $Z_{\rm el}=6\,\mathrm{kpc}$, and adopting the normalization constant $C_{\rm norm}$ from Section~\ref{sec:EdNdE_disc}, we find significantly higher amplitudes. Considering only the \textit{WMAP}–\textit{Dragon} comparison for the \textit{KST24 full} and \textit{KST24 galaxy} models, the fitted amplitudes for the different \textit{WMAP} variants differ on average by a factor of 3–4 from the \textit{Dragon} model. The reason the \textit{WMAP} model has a higher fitted amplitude is that we use a fixed value of the normalization $C_{\rm{norm}}$ from Equation~\ref{Eq_WMAP_EdNdE}, motivated by \citep{Strong2007,JF12,Jansson_2012}, which results in a less bright synchrotron sky.

The correlation coefficients for both \textit{KST24 full WMAP} and \textit{KST24 full \textit{Dragon}} models are generally similar across most regions, including \textit{Q-NW, Q-SE, North, South}, and \textit{Full} sky, with discrepancies limited to the second decimal place. A couple of outliers are regions \textit{Q-NE} and \textit{Q-SW}, where significant differences in the Spearman's correlation coefficient arise. For example, in region \textit{Q-NE}, the \textit{Dragon} model exhibits a lower correlation with the data ($r = 0.36$) compared to the \textit{WMAP} model ($r = 0.52$), mostly because the \textit{WMAP} model most likely does not drop as rapidly as the \textit{Dragon} model in density at outer longitudes.

{In Appendix \ref{sec:appendixB} we show a comparison between the fitted amplitudes and correlation coefficient obtained from the \textit{UF23} models for both the \textit{WMAP} electron distribution model and \textit{Dragon}.}

\subsection{Effect of turbulence and striated fields}

The generation of anisotropic random magnetic fields can be attributed either to magnetohydrodynamic turbulence \citep{Goldreich} or to the stretching or compression of an isotropic random magnetic field that is flux-frozen \citep{Laing1980}, i.e. one-dimensional fluctuations along a preferred orientation. \citet{Jaffe_2010} term these 'ordered random', while \citet{JF12,Jansson_2012} use the term 'striated'. In our analysis, only \textit{SVT22 str tur} and \textit{JF12 full str tur} have turbulent fields incorporated in their modelling, with \textit{SVT22} having extremely large levels of isotropic turbulence, which does not fit the data either in PI or in PA. With regard to \textit{JF12 full str tur}, the model performs better than \textit{SVT22} in PI and PA, as it has a combination of random turbulent fields and striated fields, but the level of fluctuation in the skymaps of the observed angles is still much higher than in these models.

The \textit{UF23} models do not have turbulent fields but do incorporate the contribution of striated fields by applying a simple, spatially independent, multiplicative factor to the coherent field such that $B_{\rm{striated}} = ( 1 + \beta) B$. The striated fields do not have any effect on the morphology but do change the fitted amplitude by a constant factor of $( 1 + \beta)$, as also shown in Figures~\ref{tab:amp_Spearman_final_8Quad} and \ref{tab:amp_Spearman_final_Big}.

\subsection{Spearman's coefficient}

For Spearman's correlation coefficient, we find that most models perform moderately throughout all regions; there are, however, a few outliers that we discuss below:
\begin{enumerate}
\item In Figure~\ref{fig:tt-kst24-Q-NE_uf23cre10-Q-SW} on the left-hand side, we plot the T–T plot for the \textit{KST24 full \textit{Dragon}} model against the radio data for all three regions \textit{Q-NE, Q-1-NE, Q-2-NE}. It is evident from the plot that the presence of two separate features results in an overall lower correlation coefficient for \textit{Q-NE}. The model correlates better with the data for the smaller regions \textit{Q-1-NE} and \textit{Q-2-NE}.

\item Regions \textit{Q-SW, Q-3-SW, Q-4-SW} also show a similar discrepancy in the correlation coefficient across all models. As an example, we show the T–T plots for \textit{UF23 cre10 striated Dragon} on the right-hand side of Figure\,\ref{fig:tt-kst24-Q-NE_uf23cre10-Q-SW}. Region \textit{Q-4-SW} shows higher correlation because both the data and the model have bright emission at the centre, whereas \textit{Q-3-SW} does not show morphological resemblance between data and model.
\end{enumerate}

The relatively high Spearman's rank correlations seen in broad sky regions (e.g. \textit{North, South, Full sky}) suggest that large-scale magnetic field structures are reasonably well captured by some models. Models that incorporate detailed local features, like \textit{KST24 full WMAP/\textit{Dragon}} and some \textit{UF23} variants, generally perform better, reinforcing the need for future GMF models to account for nearby foreground contributions to accurately reproduce the polarized synchrotron sky.

\subsection{Discrepancy in polarization angle and polarized intensity}

Figures~\ref{fig:PI_model_maps} and \ref{fig:pol_angles} reveal a marked discrepancy between models and radio data for PI and PA. A possible reason is that PI is influenced not only by the magnetic field strength but also by the distribution of cosmic-ray electrons, whereas PA predominantly reflects the geometry of the large-scale field. The line-of-sight integrated PI emission is dominated by the nearest emitting or Faraday-active structures. The distances to such local structures are poorly constrained, except for those traced by starlight polarization \citep{Panopoulou2021}.

Multi-wavelength evidence \citep[e.g.][]{Zhang_2024} further suggests that outflows launched from star-forming regions at a distance of 3–5 kpc from the GC in the Galactic disc produce both polarized ridges and X-ray-bright edges. If synchrotron-emitting structures in these outflows contribute significantly, they may distort the PI morphology relative to model predictions. In contrast, PA may still capture the global GMF orientation, since the outflow-driven fields could align with the large-scale structure, affecting only the inner longitude regions.

Taken together, the discrepancy between PI and PA likely reflects a combination of (i) local screens dominating the PI integral, (ii) outflows and disc–halo coupling at 3–5 kpc, and (iii) small-scale depolarization processes absent in existing GMF models. A further caveat is that most GMF models assume a toroidal halo field motivated by rotation-measure data, which naturally enforces broadly similar PA patterns. As a result, apparent agreement between models and data in PA may partly arise from shared assumptions rather than true physical accuracy. For example, the \textit{KST24} model (which incorporates the Local Bubble) and the \textit{JF12 full str} model (which does not) yield PA maps that resemble the observations overall, yet both diverge from the data in important details: the \textit{KST24} model incorporates the Fan region in its modelling but the PA map fails to match southern-hemisphere angles, while \textit{JF12 full str} exhibits very similar shortcomings despite having no Fan region in its modelling. This underlines that PA agreement alone is not a sufficient diagnostic without simultaneous consistency in PI.

\section{Conclusions}
\label{sec:conclusion}

New observations of the Galactic sky with C-BASS at 4.76~GHz open up an unexplored frequency domain for GMF modelling. In this work, we have compared several state-of-the-art GMF models and their combinations using full-sky polarized synchrotron emission maps derived from the C-BASS and S-PASS datasets. For all models, we adopted the \textit{WMAP} analytical expression for the non-thermal electron distribution to compute the polarized synchrotron emission. In selected cases, specifically for the \textit{KST24} and \textit{UF23} models, we also explored synchrotron maps generated using the non-thermal electron distribution from the \textit{Dragon} code.

Despite differences in amplitude, the correlation coefficients between the synchrotron maps generated using \textit{WMAP} and \textit{Dragon} electron distributions for \textit{KST24 full} were remarkably similar. The largest discrepancies were observed in regions \textit{Q-NE}, \textit{Q-1-SE}, and \textit{Q-SW}, the causes of which are not entirely clear.

In terms of polarization angles, the \textit{KST24 full} model and a subset of the \textit{UF23} models, particularly \textit{UF23 nebCor} and \textit{UF23 base}, show the best agreement with the data, as indicated by the lowest standard deviation in angular differences. Notably, while the polarization angle map from the \textit{LogSpiral} model visually resembles the data, its polarized intensity distribution exhibits no morphological or brightness similarity to observations and was therefore excluded from the intensity analysis.

An examination of the structure functions of polarization angles for a few selected models reveals that, in the NPS region, model predictions are more consistent with the data even though none of the models explicitly include the NPS in their construction. In contrast, at high Galactic latitudes, the observed data decorrelate more rapidly than predicted by the models. This discrepancy is likely due to the absence of turbulent magnetic fields in the model prescriptions.

Overall, we find that none of the current GMF models can simultaneously reproduce both the polarized synchrotron intensity and polarization angles; both of these observed quantities need to be fitted by the models in order to describe the GMF of the Galaxy. This likely reflects the limited treatment of local foreground structures and small-scale magnetic field variations. To advance our understanding of the GMF, future efforts must incorporate forthcoming datasets such as starlight polarization measurements from PASIPHAE \citep{pasiphae,Gina_2021} and 3-D rotation measure information from the GMIMS and POSSUM surveys \citep{POSSSUM_pol,Wolleben2009,sun2025,Ordog2025}. Additionally, with the anticipated completion of AugerPrime, the capability to identify the composition of ultra-high-energy cosmic rays on an event-by-event basis will enable rigidity-dependent studies of cosmic ray deflections \citep{Auger_Prime_2016,Auger_Prime_2019}. These advances will be crucial for refining models of the large-scale structure of the GMF.

\section*{Acknowledgments}

VS would like to thank Michael Unger and Alexander Korochkin for discussing their models and providing the Stokes \textit{Q} and \textit{U} maps. VS would also like to thank Andrew Taylor and Arjen Van Vliet for useful feedback. VS/CD/JPL/SEH acknowledge funding from the STFC (Consolidated Grant ST/P000649/1) and CD/SEH from UKSA (LiteBIRD UK ST/Y005945/1). GAH acknowledges support from the Dean’s Doctoral Scholarship at the University of Manchester. This paper uses pre-publication data from the C-BASS project, which is a collaboration between Oxford and Manchester Universities in the U.K., the California Institute of Technology in the U.S., Rhodes University, UKZN and the South African Radio Astronomy Observatory in South Africa, and the King Abdulaziz City for Science and Technology (KACST) in Saudi Arabia. The work at Oxford was supported by funding from STFC, the Royal Society, and the University of Oxford. The work at the California Institute of Technology and Owens Valley Radio Observatory was supported by National Science Foundation (NSF) awards AST-0607857, AST-1010024, AST-1212217, and AST-1616227, and by NASA award NNX15AF06G. The work at Manchester was supported by several STFC Consolidated Grants and a UKSA grant (ST/Y005945/1) funding LiteBIRD foreground activities. A.C. Taylor, M.E. Jones, and G. Weymann-Despres also acknowledge support from the Horizon Europe project RadioForegroundsPlus (GA 101135036), which is supported in the U.K. by UKRI grant number 10101603.

\section*{Data availability}
We make use of the HEALPix~\citep{Healpix_2005}, Healpy~\citep{Healpy_2019}, Matplotlib~\citep{Matplotlib} and Numpy~\citep{Numpy} packages. The codes used for this work can be made available to the corresponding author upon request. 

\bibliographystyle{mnras}
% Use the LaTeX power, use bibtex properly.
% \bibliography{references}
\bibliography{sorted_references,cbassbiblio}

\appendix

{\section{Effect of different electron-distribution for \textit{UF23} models}
\label{sec:appendixB}

In Figure~\ref{fig:uf23_compare_ednde}, we plot the fitted amplitudes and correlation coefficients across regions for all \textit{UF23} variants, with the \textit{Dragon} model in red and the \textit{WMAP} model in black. Similar to \textit{KST24}, in the \textit{UF23} models discrepancies between \textit{WMAP} and \textit{Dragon} versions lie mostly in the fitted amplitude. The fitted amplitudes from \textit{UF23 WMAP} variants are 8–10 times larger than the \textit{Dragon} variants. The Spearman's correlation coefficients calculated from the \textit{WMAP} and \textit{Dragon} versions for the respective \textit{UF23} models have similar values, thereby suggesting that these two electron distributions do not introduce significant morphological differences in the polarized intensity maps but drastically impact the net overall synchrotron brightness.
\begin{figure*}

   \includegraphics[width=1.0\linewidth]{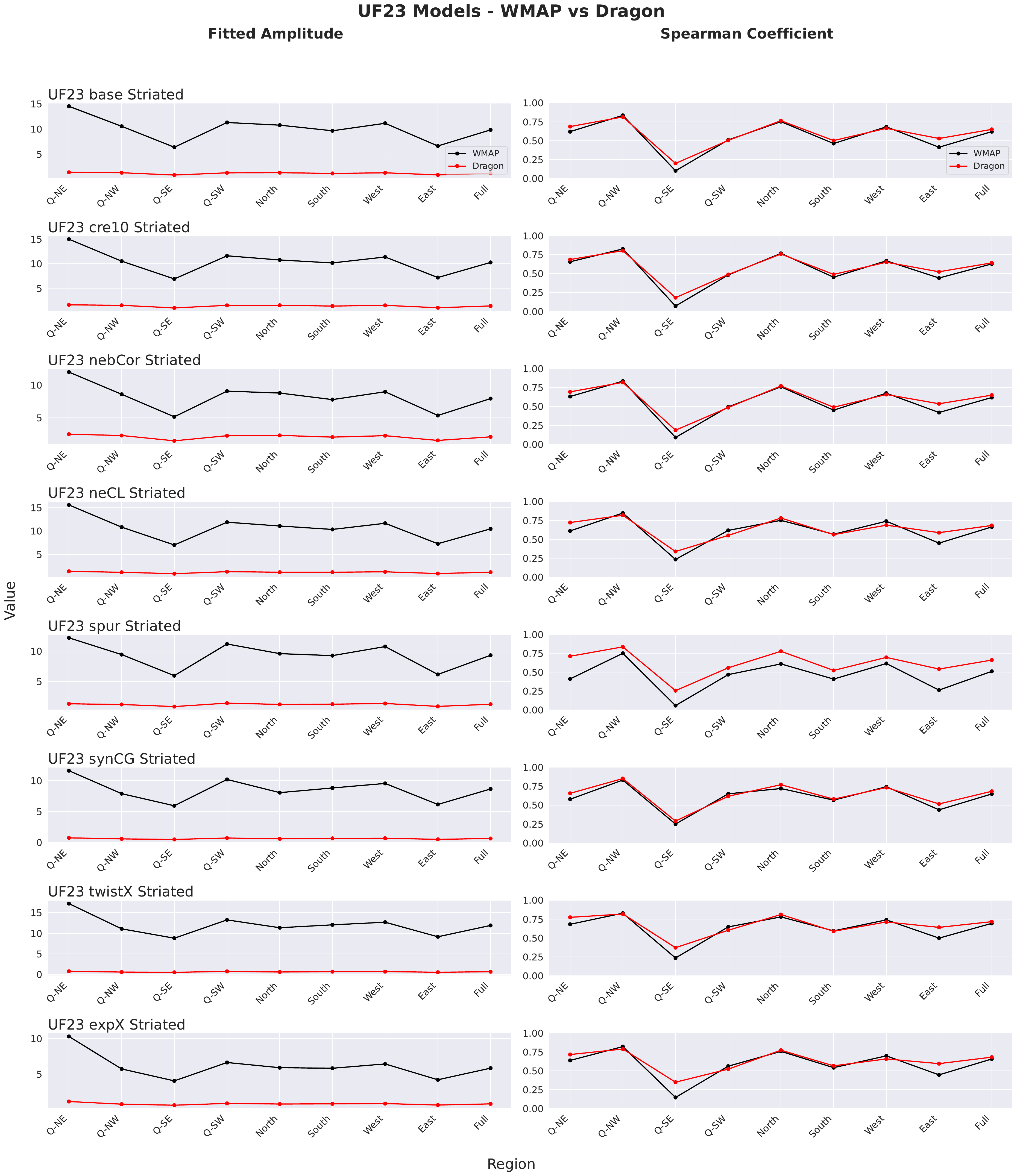}

    \caption{Fitted amplitudes and Spearman's coefficient for \textit{UF23} models for different electron distributions in different regions of the same sky. We compare the different \textit{UF23} model for the \textit{\textit{Dragon}} model and the \textit{WMAP} model, with radial and azimuthal cut-off of 5\,kpc and 6\,kpc respectively.}
    \label{fig:uf23_compare_ednde}
\end{figure*}}

% \end{figure*}
{
\section{Comparison between C-BASS and S-PASS in the overlap region}
\label{sec:appendixC}

\begin{figure*}
    \centering
    \includegraphics[width=1.0\linewidth]{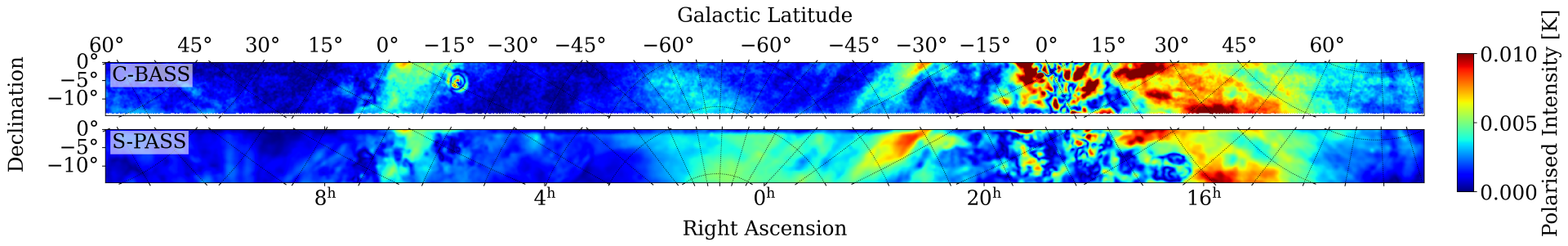}
    \caption{Polarized intensity in the overlap declination strip ($-15^\circ < \delta < -1^\circ$) between C-BASS (above) and S-PASS (below). The S-PASS map has been scaled to 4.76\,GHz and smoothed to $1^\circ$ resolution as described in Sec.~\ref{sec:masks}. Grid lines show Galactic coordinates.}
    \label{fig:strip}
\end{figure*}

To assess the validity of the S-PASS scaling, we compare the scaled 2.3\,GHz map with the C-BASS 
4.76\,GHz data in the declination overlap region ($-15^{\circ} < \delta < -1^{\circ}$), where both 
surveys provide independent measurements (Fig.\,\ref{fig:strip}). Restricting the analysis to this 
common sky area allows a direct empirical test of the adopted power-law extrapolation. While 
the two datasets are consistent over much of this region, morphological discrepancies are 
present in certain areas, as discussed below. We 
have chosen to use the C-BASS map down to $-10^{\circ}$ since it is at our target frequency and also suffers from less Faraday rotation than S-PASS.
A detailed analysis of the region of largest disagreement, including comparison with several other sky surveys, will be presented in a companion paper (Cepeda-Arroita et al. in preparation).

The scaled S-PASS map is in good agreement with C-BASS (see Figure \ref{fig:strip}) where the signal is strong, with the 
exception of regions close to the Galactic plane, where enhanced Faraday depolarization 
at 2.3\,GHz is evident. This depolarization extends to $b = 30^{\circ}$ near RA\,16$^{\rm h}$30$^{\rm m}$, 
attributable to Sh\,2-27, a known source of large rotation measure gradients 
\citep{2022A&A...663A.170R, 2019MNRAS.487.4751T}. The most pronounced disagreement elsewhere is 
in the high-latitude southern Galactic hemisphere near RA\,$0^{\rm h}$, where the polarized emission 
is generally faint, but S-PASS shows notably more of it than C-BASS. Additionally, leakage of total 
intensity emission from the bright, nearly unresolved Orion Nebula (J0535$-$05) is visible in the 
C-BASS map.

We show T-T plots of the overlap region split by Galactic latitude, with and without the RM 
correction mask applied, in Figure.\,\ref{fig:Overlap} with the region of $b > 0^{\circ}$ shown on the left-hand side and $b < 0^{\circ}$ shown on the right-hand side of the figure.

For $b > 0^{\circ}$ (Figure\,\ref{fig:Overlap}, see left), the overlap region is dominated by the 
North Polar Spur (NPS), clearly visible in both maps in Fig.\,\ref{fig:strip}. Without the RM 
mask the two datasets already show reasonable agreement ($r = 0.78$, $N = 104$); applying the 
mask improves this substantially to $r = 0.96$ ($N = 63$), confirming both that the NPS is 
consistently recovered in both surveys and that the RM masking procedure effectively removes 
pixels with significant Faraday contamination. For $b \leq 0^{\circ}$ 
(Figure\,\ref{fig:Overlap}, see right), the correlation is poor, the slope is shallow, although it would become steeper if the relation was forced to go through the origin. The correlation is slightly improved by masking 
($r = 0.32$, $N = 119$ without; $r = 0.41$, $N = 75$ with); the slope becomes marginally flatter, though this change is likely within the errors.

\begin{figure*}
    \centering
    \includegraphics[width=0.49\linewidth]{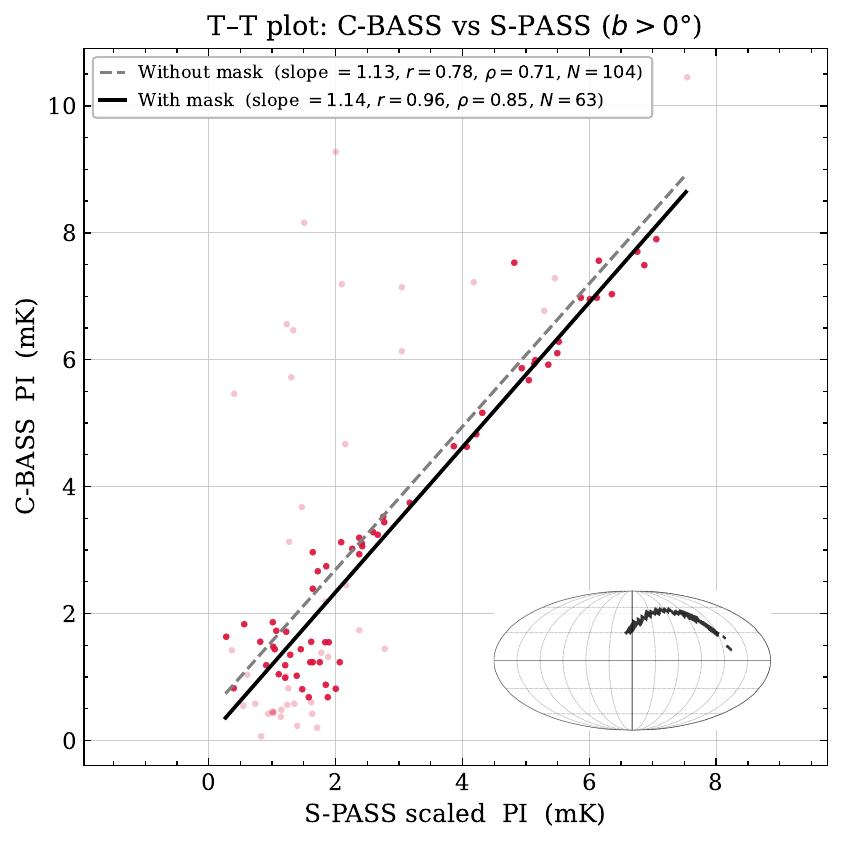}
    \includegraphics[width=0.49\linewidth]{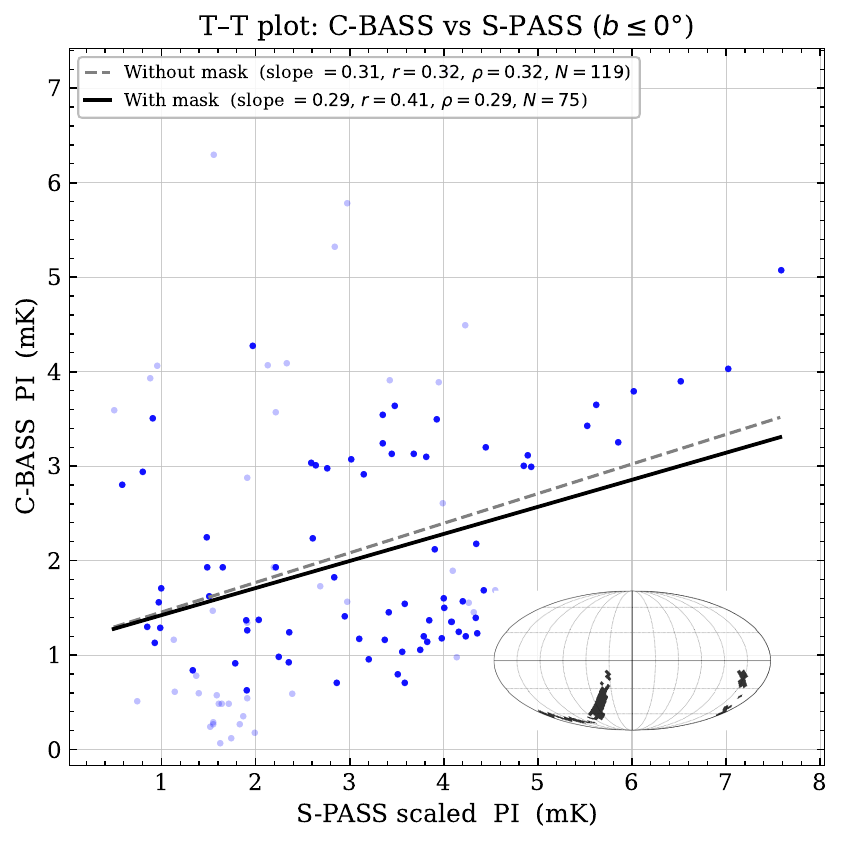}
    \caption{{\textbf{Left:} T--T plot of C-BASS versus scaled S-PASS polarized intensity for the 
    overlap region at $b > 0^\circ$, dominated by the North Polar Spur (NPS). 
    Faint points show all pixels in the full unmasked overlap strip 
    ($-15^\circ < \delta < -1^\circ$, $N = 104$, Pearson $r = 0.78$, 
    Spearman $\rho = 0.71$, slope $= 1.13$); solid points show only those 
    retained after the RM correction mask is applied ($N = 63$, Pearson $r = 0.96$, 
    Spearman $\rho = 0.85$, slope $= 1.14$). The masked slope is consistent with 
    the expected spectral scaling between the two frequencies. The improvement in 
    correlation upon masking indicates that residual Faraday rotation in the excluded 
    pixels decorrelates the polarized emission between C-BASS and S-PASS. The inset 
    Mollweide map shows only the pixels retained after masking; the full unmasked strip 
    runs down to $b = 0^{\circ}$ at both ends.
    \newline \textbf{Right}:T--T plot of C-BASS versus scaled S-PASS polarized intensity for the 
    overlap region at $b \leq 0^\circ$. Faint points show all pixels in the full 
    unmasked overlap strip ($-15^\circ < \delta < -1^\circ$, $N = 119$, Pearson 
    $r = 0.32$, Spearman $\rho = 0.32$, slope $= 0.31$); solid points show 
    only those retained after the RM correction mask is applied ($N = 75$, Pearson 
    $r = 0.41$, Spearman $\rho = 0.29$, slope $= 0.29$). The correlation remains poor regardless of masking, suggesting the discrepancy is driven by systematic effects rather than Faraday rotation. The inset Mollweide map shows only the 
    pixels retained after masking; the full unmasked strip runs up to $b = 0^{\circ}$ 
    at both ends.}}
    % \label{fig:Overlap_b_gt_0}
    \label{fig:Overlap}

\end{figure*}

}
s

\end{document}